\documentclass[%
reprint,
amsmath,amssymb,
aps,
]{revtex4-2}

\usepackage{graphicx}
\usepackage{dcolumn}
\usepackage{bm}
\usepackage{hyperref}
\usepackage[mathlines]{lineno}

\begin{document}
	
	\preprint{APS/123-QED}

\title{Domain-Growth Kinetics and Scaling Laws Governing Pulse-Driven Accumulative Polarization Switching in HZO}
\author{Manish Anand}
\email{itsanand121@gmail.com}
\thanks{Corresponding Author}
\affiliation{Department of Physics, Bihar National College, Patna University, Patna-800004, India.}
\author{Balram Khattar}
\email{balram.22eez0016@iitrpr.ac.in}
\affiliation{Department of Electrical Engineering, Indian Institute of Technology Ropar, Rupnagar, 140001, India}

\author{Abhishek Sharma}
\email{abhi@iitmandi.ac.in}
\affiliation{School of Computing and Electrical Engineering, Indian Institute of Technology Mandi, Mandi, 175005, India}
\date{\today}

\begin{abstract}
Accumulative polarization switching driven by sequential sub-coercive electric-field pulses offers a promising route toward low-power ferroelectric memories and neuromorphic devices. However, the kinetic regimes governing this nonequilibrium process remain poorly understood. Here, we employ a phase-field model based on the time-dependent Landau–Ginzburg formalism to investigate pulse-driven accumulative switching in ferroelectric HZO. By systematically varying the initial domain configuration, pulse amplitude, pulse-on time, and pulse-off time, we establish a quantitative link between microscopic domain-wall dynamics and macroscopic polarization accumulation. We show that the effective switched-domain radius, (R), follows distinct scaling regimes characterized by the local kinetic exponent, $\alpha^{}_{\rm local}=d\ln(R-R^{}_0)/d\ln n$. Initially, $\alpha^{}_{\rm local}>1$ indicates superlinear domain growth driven by enhanced irreversible domain-wall propagation under successive pulses. As switching progresses, $\alpha^{}_{\rm local}\approx1$ marks steady self-similar growth, whereas $\alpha^{}_{\rm local}<1$ signifies decelerating dynamics caused by geometric confinement, depletion of switchable polarization, and relaxation-induced back switching. The transition between these regimes is governed by the competition between field-driven excitation during the pulse-on interval and spontaneous relaxation during the pulse-off interval, and is further influenced by the initial domain geometry. Increasing the pulse amplitude or pulse-on duration extends the super-linear regime, whereas longer pulse-off times promote relaxation and suppress accumulation. These findings establish a unified scaling framework for pulse-driven accumulative switching, providing quantitative insight into nonequilibrium ferroelectric domain evolution and design guidelines for HZO-based memory and neuromorphic devices.

\end{abstract}

\maketitle
\newpage
\section{Introduction}
Ferroelectric (FE) materials are of profound importance due to their immense industrial applications~\cite{xu2008phase,segatto2025lgd,jiao2025temperature,koduru2025small,kumar2025modeling,wang2006size}. In particular, Zr-doped HfO$^{}_2$
(Hf$_{1–x}$Zr$_{x}$O$_{2}$:HZO) have received significant attention due to its wide range of technologically relevant properties~\cite{gaddam2022novel,chen2024impact,yu2023interfacial,yoo2023recent,wu2025review}. For instance, it has got thickness scalability, CMOS process compatibility, high endurance, low switching energy, and many promising attributes of ferroelectric field effect transistors (FEFETs), etc~\cite{qin2024hzo,jan2023operando}. Due to these, HZO is currently the most important ferroelectric material for CMOS-compatible ferroelectric memories and logic devices~\cite{khattar2025asymmetric}. Furthermore, the recently reported accumulation-mediated polarization switching process under the application of a sub-coercive field pulse train enables low-power, reliable, and analogue control of ferroelectric states through progressive domain nucleation and growth, making it highly attractive for multi-level memories, neuromorphic computing, and energy-efficient ferroelectric devices~\cite{saha2019phase,ni2018memory,mulaosmanovic2018}.

Accumulative switching enables polarization reversal through the repeated application of sub-coercive electric-field pulses, each of which is individually insufficient to induce complete switching~\cite{gao2025reconfigurable,jiang2026ferroelectric}. By permitting progressive domain evolution and net polarization build-up under such conditions, this mechanism extends the operating window of ferroelectric devices into the sub-coercive-field regime. Besides, such polarization switching schemes also provide  low-voltage and energy-efficient operation compared with conventional single-pulse switching and hence provide access to stable intermediate polarization states, which improve device reliability, and support analogue functionalities suitable for multi-level memory and neuromorphic computing applications~\cite{mulaosmanovic2018mimicking,dutta2020supervised}.  In these contexts, a few works are worth mentioning. Using computer simulations and experiments,  Atanu {\it et al.} demonstrated the accumulative polarization switching (P-switching) in FE thin films (HZO) under the influence of sequential sub-coercive electric-field pulses~\cite{saha2019phase}. They observed that the domain wall (DW) undergoes spontaneous motion even in the absence of an applied electric field. Such DW instability leads to spontaneous P-excitation and relaxation processes, which play a pivotal role in accumulative P-switching in an FE grain. Mulaosmanovic {\it et al.} investigated the accumulative switching property of ferroelectric HfO$^{}_2$ embedded within a nanoscale FeFET device~\cite{mulaosmanovic2018}. They observed that the device undergoes a complete and abrupt switching in response to a train of identical incident gate-voltage pulses, even though an individual voltage pulse is insufficient in magnitude for switching. Such a phenomenon could be attributed to a progressive nucleation of ferroelectric domains, which gain in size and quantity with an increase  in the number of incoming pulses. Mulaosmanovic {\it et al.} experimentally investigated accumulative switching and one-shot switching in large and small-area FeFETs~\cite{mulaosmanovic2020}. They showed that both switching types obey the universal time-voltage dependence, irrespective of the device size, pulse polarity, or time intervals between the excitation pulses. Changhyeon {\it et al.} studied the tunable ferroelectric properties of HZO, particularly focusing on
coercive voltage ($V^{}_{CO}$) and remnant polarization ($P^{}_{CO}$) through controlled cycling voltage~\cite{han2025tunable}. They found a  crucial relationship between the ferroelectric properties and the proportion of the ferroelectric phase.

FE systems are also emerging as one of the most promising candidates for neuromorphic computing and logic-in-memory (LiM) architectures, as polarization dynamics in such a system (in response to external stimuli, for instance, an electric field) closely emulate biological synapses, a feature crucial for learning and adaptation in neural networks~\cite{xue2021giant,xiao2026multi,yoo2026reconfigurable}. For efficient usage of FE systems for these applications requires multiple intermediate states between fully polarized states, which can be achieved by application of a voltage pulse train, an efficient en route to polarization accumulation. More importantly, such gradual polarization switching can be efficiently exploited for weight modulation of synapses in FE-based neuromorphic systems. Consequently, sequential voltage pulse trains could transform a ferroelectric device from a binary memory element into an analogue adaptive synapse by exploiting accumulative polarization
switching~\cite{shen2026multilevel,sunbul2021optimizing}. The resulting gradual, non-volatile, and energy-efficient modulation of polarization enables the emulation of biological learning processes such as Long-Term Potentiation (LTP), Long-Term Depression (LTD), and spike-dependent plasticity, making ferroelectrics highly attractive for neuromorphic computing applications~\cite{max2020hafnia,oh2019ferroelectric}. Wang {\it et al.} showed that accumulative-mediated polarization switching enables high-resolution analogue synapses due to gradual domain-wall motion~\cite{wang2026manipulating}. Gao {\it et al.} found that the weight updates in FE-based neuromorphic systems can be accomplished from cumulative polarization processes~\cite{gao2025reconfigurable}. Jadhav {\it et al.} addressed challenges in implementing non-volatile, analogue synaptic elements for neuromorphic hardware~\cite{jadhav2026lorentzian}. By leveraging Lorentzian switching dynamics, they aimed to improve linearity and symmetry in synaptic weight updates owing to partial domain switching. Lee {\it et al.} investigated the fabrication of high-performance HZO-based synaptic devices by controlling the concentration of oxygen vacancies $V^{}_O$ and tuning the corresponding ferroelectric properties of the HZO active layer~\cite{lee2025polarization}. They showed that the linearity and symmetry of synaptic weight modulations could be effectively improved by controlling pulse widths and pulse intervals. Xiang {\it et al.} successfully fabricated a multigate
two-dimensional semiconductor FeFET, a potential candidate for neuromorphic and LiM computing applications~\cite{xiang2025ferroelectric}. They achieved such multi-functionality by controlling ferroelectric polarization precisely and hence leveraging intermediate polarization states obtained through phase engineering of HZO films. These studies clearly show that the accumulation-mediated polarization switching process enables gradual and controllable modulation of ferroelectric polarization through repeated electrical stimuli, thereby emulating synaptic plasticity in biological neural networks, making it an ideal system with multi-level memory functionalities suitable for neuromorphic computing systems.


The above discussion clearly suggests that accumulation-mediated polarization switching induced by sequential voltage pulse trains offers significant potential for a wide range of emerging technological applications, particularly neuromorphic computing, where gradual and history-dependent polarization evolution is essential for synaptic functionality. This phenomenon is also of considerable fundamental interest, as it provides valuable insights into the nonequilibrium kinetics of polarization evolution, domain-growth dynamics, and the interplay between switching and relaxation processes under repetitive excitation inducsed by field pulse train. Although both experimental and theoretical studies on switching through polarization accumulation in HZO have been undertaken. They primarily focused on understanding the physical origin of polarization accumulation in HZO through domain-wall motion and spontaneous excitation/relaxation processes leading to gradual polarization buildup under repeated sub-coercive pulses~\cite{saha2019phase}. In the context of neuromorphic-based applications of FE systems, the above-mentioned studies also clearly clarify the fact that in practical ferroelectric-based neuromorphic devices, the polarization state is not established in a single step but evolves through repeated sub-threshold voltage pulses, where partial switching, domain nucleation, growth, and relaxation compete on comparable timescales. However, the underlying kinetic regimes governing this accumulative switching, especially the interplay between pulse width, inter-pulse delay, and relaxation-driven back-switching, are still not fully understood. As a result, there is a lack of predictive scaling laws that can connect microscopic domain evolution to macroscopic polarization build-up under realistic driving conditions. Therefore, there is a pressing need for a systematic investigation to identify the kinetic regimes of polarization accumulation, establish the scaling laws governing pulse-driven switching, and elucidate how domain-growth and relaxation processes collectively determine the final polarization state. Such understanding is essential not only for reliable ferroelectric memory operation but also for optimizing HZO-based neuromorphic elements where gradual, repeatable, and controllable weight update behaviour is required.
	
Thus motivated, in the present work, we go beyond conventional studies of polarization accumulation by uncovering the domain-growth kinetics and scaling laws governing pulse-driven switching in HZO. Through the combined analysis of domain morphology, domain radius evolution, local scaling exponents, and pulse-parameter dependence, we attempt to establish a unified framework linking accumulative switching to the competition between field-assisted activation (electric field $E^{\rm app}_{n}$, Pulse-on time/pulse width $T^{}_{\mathrm {on}}$) and Pulse-off time/relaxation time ($T^{}_{\mathrm {off}}$). Such identification of distinct dynamic switching regimes and scaling behaviour provides fundamental insights into ferroelectric memory and neuromorphic device operation that are largely absent in existing HZO literature.
\section{Model}
 In the present work, we have used a dynamic phase field model to investigate polarization accumulation in FE systems with diverse classes of initial domain configurations upon application of a voltage pulse train. We further investigate domain-growth kinetics and underlying scaling laws which govern pulse-driven polarization switching. The spatial and temporal evolution of polarization (P) can be calculated using the time-dependent Landau-Ginzburg (TDLG) equation~\cite{hoffmann2018ferroelectric,bandyopadhyay2006approach}

\begin{equation}
-\rho\frac{\partial\vec{P}}{\partial t}=\frac{\delta F}{\delta\vec{P}}
\end{equation}
Here, $\rho$ is the viscosity coefficient, and $t$ is the time. $F$ is the free energy of the system defined as~\cite{nambu1994domain,rabe2007physics} 
\begin{equation}
	F=\int \Big [- \frac{K^{}_{P}}{2}|\nabla\vec{P}|^2+\frac{\alpha}{2}|\vec{P}|^2+\frac{\beta}{4}|\vec{P}|^4+\frac{\gamma}{6}|\vec{P}|^6-\vec{E}^{app}\cdot\vec{P} \Big ]dV
\end{equation}
Here, the first term is the gradient energy, which controls the motion and interaction of domain walls, thereby influencing domain growth kinetics. The $2^{\rm nd}$ to $4^{\rm th}$ terms represent Landau energy, which describes the intrinsic thermodynamic stability of the ferroelectric phase and hence determines polarization states of the underlying system. The last term denotes electrostatic energy due to an interaction between polarization and the externally applied electric field.

We assume the polarization direction along the thickness ($z$-axis) of the FE, i.e. $P^{}_{x}=0$, $P^{}_{y}=0$, $P^{}_z\neq0$ \cite{o2018stabilization,materlik2015origin}. Therefore, the above equation takes the following form~\cite{nambu1994domain,rabe2007physics}:
\begin{equation}
	-\rho\frac{d P^{}_z}{d t}=\frac{\delta F}{\delta P^{}_z}
\end{equation}

\begin{equation}
-\rho\frac{d P^{}_z}{d t}=-K^{}_{P}\nabla^2 P^{}_z+{\alpha}P^{}_z+{\beta}P^{3}_{z}+{\gamma}P^{5}_{z}-E^{app}_{z}
\label{TDGL1}
\end{equation}
Normalizing the applied electric field $E^{app}_{z}$ and polarization $P^{}_z$ with respect to $E^{}_{C0}$ and $P^{}_{C0}$, respectively. Here $E^{}_{C0}$ is the coercive field (the electric field value at which $\frac{dE^{\rm app}_{z}}{dP^{}_{z}}=0$) of a non-interacting
lattice, and $P^{}_{C0}$ is the value of polarization at the electric field $E=E^{}_{C0}$~\cite{saha2019phase}. We can then write the equation.~(\ref{TDGL1}) as
\begin{widetext}
\begin{equation}
-\frac{\rho P^{}_{C0}}{E^{}_{C0}}\frac{d\bigg(\frac{P^{}_z}{P^{}_{C0}}\bigg)}{dt}=-\frac{E^{app}_{z}}{E^{}_{C0}}-\frac{K^{}_{P}P^{}_{C0}}{E^{}_{C0}}\nabla^{2}\bigg(\frac{P^{}_z}{P^{}_{C0}}\bigg)+\frac{\alpha P^{}_{C0}}{E^{}_{C0}}\frac{P^{}_z}{P^{}_{C0}}+\frac{\beta P^{3}_{C0}}{E^{}_{C0}}\bigg(\frac{P^{}_z}{P^{}_{C0}}\bigg)^3+\frac{\gamma P^{5}_{C0}}{E^{}_{C0}}\bigg(\frac{P^{}_z}{P^{}_{C0}}\bigg)^5
\end{equation}
\end{widetext}
The above equation can then be written as~\cite{saha2019phase}
\begin{equation}
-\rho_{n}\frac{\partial P^{}_n}{\partial t}=-E^{app}_{n}-K^{n}_{P}\nabla^2 P+\hat{\alpha}P^{}_n+\hat{\beta}P^{3}_{n}+\hat{\gamma}P^{5}_{n}
\label{TDGL}
\end{equation}

Here, $P^{}_n=P^{}_{z}/P^{}_{C0}$, $E^{app}_{n}=E^{app}_{z}/E^{}_{C0}$, $\rho^{}_{n}=\frac{\rho P^{}_{C0}}{E^{}_{C0}}$, $K^{n}_{P}=\frac{K^{}_{p}P^{}_{C0}}{E^{}_{C0}}$, $\hat{\alpha}=\frac{\alpha P^{}_{C0}}{E^{}_{C0}}$, $\hat{\beta}=\frac{\beta P^{3}_{C0}}{E^{}_{C0}}$, and $\hat{\gamma}=\frac{\beta P^{5}_{C0}}{E^{}_{C0}}$. Here, $P^{}_n$ and $E^{app}_{n}$ are the polarization and the applied electric field normalized to $P^{}_{C0}$ and $E^{}_{C0}$, respectively. $\rho_{n}$ is the normalized kinetic coefficient and $K^{n}_{P}$ is the normalized domain-interaction parameter equivalent to the gradient energy coefficient. $\hat{\alpha}$, $\hat{\beta}$, and $\hat{\gamma}$ are the normalized effective Landau coefficients with their calibrated values -1.499, +0.498, and 0.001, respectively~\cite{saha2019phase}.

We have considered an FE system of size 80 ${\mathrm{nm}} \times$ 80 ${\mathrm{nm}}$ to provide sufficient space for the evolution and interaction of multiple ferroelectric domains while maintaining computational efficiency for long-time pulse-driven phase-field simulations. We have also performed test simulations with various lateral dimensions and observed qualitatively similar polarization accumulation behaviour and switching kinetics. We performed finite element-based simulations to solve Eq.~(\ref{TDGL}) in a self-consistent manner on a real space grid by imposing the Neumann
boundary condition at the edges of the system~\cite{saha2019phase,cano2010multidomain}. It is important to mention that the $P^{}_n$  denotes normalized microscopic polarization in each grid point, while the analogous quantity of experimentally measured $P$ is the spatial average of $P^{}_{n}$, denoted as $\bar{P}^{}_{n}$. In the present model, the local effective interaction  field $E^{n}_{\rm int}$ is provided by the domain interaction term $(K^{n}_{P}\nabla^2 P)$. Therefore, the complete polarization reversal ($P-$switching) is dictated by $E^{\rm app}_{n}+E^{int}_{n}$. For
instance, P-switching will occur for $|E^{\rm app}_{n}+E^{n}_{\rm int}|>1$~\cite{saha2019phase}.

\begin{figure}[!htb]
	\centering\includegraphics[scale=.18]{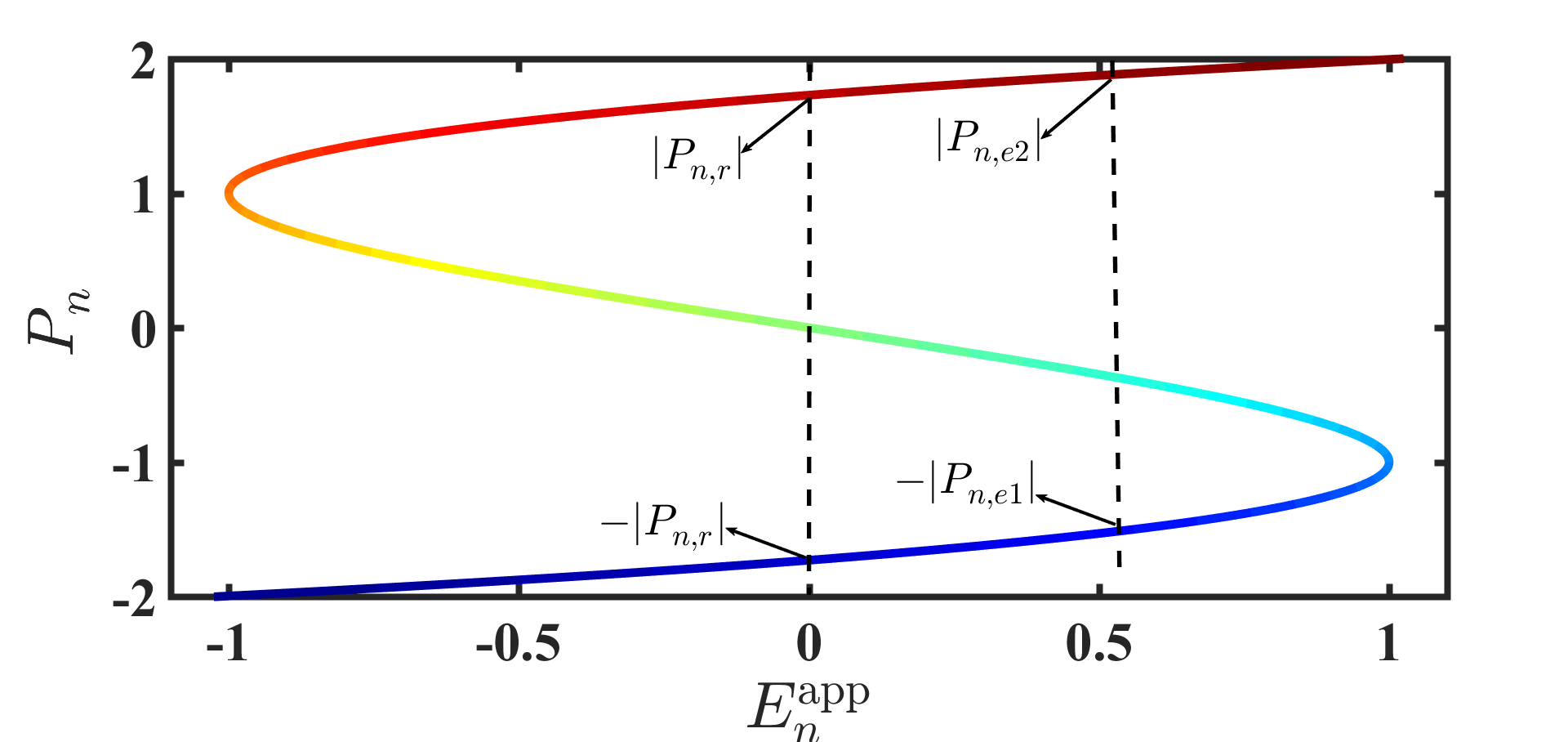}
	\caption{The variation of $P^{}_{n}$ as a function of $E^{\rm app}_{n}$. We have also shown various possible routes for excitation/relaxation (type-I, type-II. and type-III).}
	\label{Figure1}
\end{figure}

In the case of a voltage pulse train of sub-coercive field strength, the polarization switching does not occur abruptly; rather, there is a gradual build-up of polarization which eventually leads to switching. During such a process, there may be an increase or decrease in microscopic polarization value from its remanence $P^{}_{n,r}$, termed as  $P-$excitation or $P-$relaxation, respectively. In such a case, polarization can change its value through various routes. For instance, when the electric field is on, polarization can increase its value either from (i)-$|P^{}_{n,r}|$ to -$|P^{}_{n,e1}|$ [see Fig.~(\ref{Figure1})] or (ii) +$|P^{}_{n,r}|$ to +$|P^{}_{n,e2}|$[see Fig.~(\ref{Figure1})], termed as type-I and type-II excitation, respectively. On the other hand, the polarization (decreases) returns to its remanent value either from (i)-$|P^{}_{n,e1}|$ to -$|P^{}_{n,r}|$ or (ii)+$|P^{}_{n,e2}|$ to +$|P^{}_{n,r}|$ in the absence of an external electric field, referred to as type-I and type-II -relaxation, respectively. These phenomena can be understood as soft dielectric-type capacitive charging/discharging events, driven by  the electric field~\cite{eliseev2012domain}. 
Here, soft dielectricity denotes an interesting property of FE materials in which the change in dipole moment $(dP/dE)$ due to an applied E-field without any polarization switching.
It is quite clear from the above discussion that Type-I and Type-II excitation/relaxation processes generally occur under sub-coercive fields. Due to small strength, the field pulses produce partial switching and polarization accumulation rather than complete polarization reversal. Such gradual polarization accumulation leads to complete switching if the polarization increased during each pulse is greater than the polarization decreased during the off-duration of the pulse. As the external field pulse train continues, the switched domains continue to grow, eventually leading to complete switching. Therefore, such an excitation process where the polarization value increases from -$|P^{}_{n,e1}|$ to $|P^{}_{n,e2}|$ is termed as Type-III excitation [see Fig.~(\ref{Figure1})]~\cite{saha2019phase}. Interestingly, such an excitation process can also be initiated at a relatively earlier stage in the system with an initial pinned-domain near the edge. In such a case, fewer electric field pulses are required for the domain to span the ferroelectric layer, leading to an earlier transition from accumulative domain growth (Type-II) to complete polarization switching (Type-III). This behaviour is reflected by a larger initial local exponent and a more rapid increase in the domain radius compared with an equivalent domain initially located at the centre of the system.

To quantify the various class of excitation/relaxation process in such a system, the evolution of the domain radius $R$ could provide a direct microscopic measure of the competition between field-driven domain-wall motion during the pulse-on period and relaxation during the pulse-off period. The study of time evolution of $R$ could further help us in understanding the effect of various parameters of interest such as field strength, pulse-on and off time duration, initial domain configurations on domain evolution kinetics and hence underlying scaling laws. 

The evolution of the domain radius is expected to follow a power-law scaling with pulse number, $R\sim n^\alpha$, during a self-similar growth regime, $\alpha$ being the exponent and $n$ refers to pulse number~\cite{bray1994theory,grest1984kinetics}. However, because pulse-driven accumulative switching involves continuous changes in the competition between field-driven excitation and relaxation, a single global exponent cannot adequately describe the switching kinetics. Therefore, following the concept of local scaling exponents used in dynamic scaling and domain-growth kinetics, we define the instantaneous local exponent $\alpha^{}_{\rm local}$ as~\cite{barabasi1995fractal}
\begin{equation}
	\alpha^{}_{\rm local}=\frac{d \log(R-R^{}_{0})}{d \log {n}}
\end{equation}
where $R$ is the average domain radius after the $n^{\rm th}$ pulse and $R^{}_0$ is the initial domain radius. The local exponent provides a direct measure of the evolution of domain-wall propagation during accumulative polarization switching. The evolution of $\alpha_{\mathrm{local}}$ therefore could provide a direct kinetic signature of the excitation/relaxation regimes in accumulation-mediated polarization switching processes.

\section{Results}
\begin{figure*}
	\includegraphics[scale=.65]{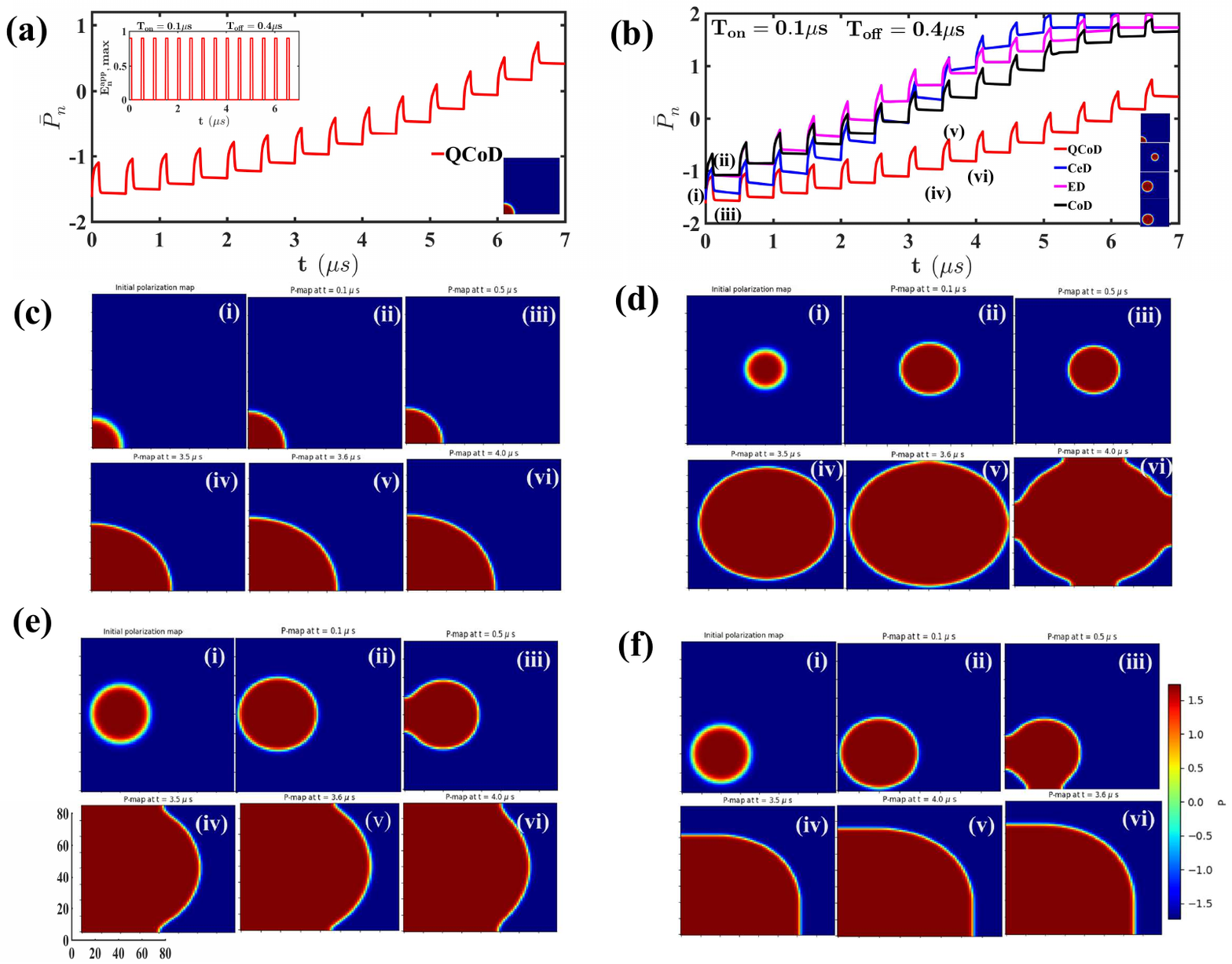}
	\caption{Pulse-driven accumulative polarization switching in HZO for four distinct domain configurations: quasi-centre domain (QCoD), centre domain (CeD), edge domain (ED), and corner domain (CoD). (a) The temporal evolution of average transient polarization $\bar{P}_{n}$ for the QCoD configuration; inset shows the sequential voltage pulse train employed in the phase-field simulations ($T^{}_{\rm on}=0.1 \mu$ s and $T^{}_{\rm off}=0.4 \mu$. (b) Evolution of the normalized average polarization for different initial domain morphologies under identical electrical excitation. (c–f) Temporal evolution of the polarization morphology for representative initial domain configurations at typical instants of time. The distinct growth pathways suggest that the initial domain geometry governs the balance between excitation and relaxation processes, resulting in different polarization accumulation rates despite converging to a common saturated state.}
	\label{Figure2}
\end{figure*}

In Fig.~(\ref{Figure2}), we study the polarization accumulation characteristics and associated excitation/relaxation features of four representative ferroelectric systems containing initially pinned circular domains located at different positions, namely the quarter-corner domain (QCoD), centrally located domain (CeD), edge domain (ED), and corner domain (CoD). We have applied an electric field of amplitude $E^{\rm app}_{n}, {\rm max}=0.8$, with $T^{}_{\rm on}=0.1$ $\mu s$ and $T^{}_{\rm off}=0.4$ $\mu s$ in all the cases. We have plotted $\bar{P}^{}_{n}$ vs. $t$ for QCoD in Fig.~\ref{Figure2}(a); inset figure also shows the applied external field pulse train, while Fig.~\ref{Figure2}(b) has  $\bar{P}^{}_{n}$ vs. $t$ for all four cases. It is quite interesting to note that although all systems are subjected to the same external field, their polarization accumulation behaviour differs significantly, indicating a decisive role of initial domain configuration on switching kinetics. The polarization evolution also exhibits a staircase-like profile for all configurations due to a net increase in polarization despite partial relaxation during the off period. 
Another important thing to note is that $\bar{P}^{}_{n}$ reaches its maximum at the fastest rate for the CeD configuration as the circular domain wall is symmetrically surrounded by negative polarization and can therefore propagate simultaneously in all radial outward directions. The latter induces an isotropic driving force acting along the entire circumference, maximizing the effective Type-III excitation, resulting in rapid domain expansion and the highest polarization accumulation rate. On the contrary, $\bar{P}^{}_{n}$ variation is the slowest for QCoD, because the initially switched domain occupies only one corner of the system. 

In Fig.~\ref{Figure2}(c)-Fig.~\ref{Figure2}(f), we have shown domain morphologies at various typical time instants to understand the microscopic origin of the accumulation process. During each pulse, the negative polarization matrix undergoes Type-I excitation, making the local $P$  less negative even without changing its orientation, while the already switched $P$ experiences Type-II excitation, resulting in a slight increase in the positive polarization in all four systems. The type-III excitation contributes significantly to the observed polarization morphology, where the domain wall enters into the surrounding negative matrix, causing irreversible polarization reversal. Interestingly, the Type-I and Type-II relaxations tend to restore the polarization magnitude toward their remanent values during the pulse-off period. whereas, Type-III relaxation instigates a slight backward motion of the domain wall due to reduction in gradient energy. Specifically, the entire circular domain wall participates in Type-III excitation, producing the fastest switching in the CeD configuration. On the other hand,  two intersecting boundaries further restrict domain-wall propagation in the CoD and QCoD configurations, thereby suppressing Type-III excitation. Since Type-I and Type-II processes occur locally, the dominant difference among the four systems arises from the available geometry for irreversible Type-III mediated domain-wall motion, resulting in highly geometry-constrained polarization accumulation characteristics.

\begin{figure*}
	\includegraphics[scale=0.65]{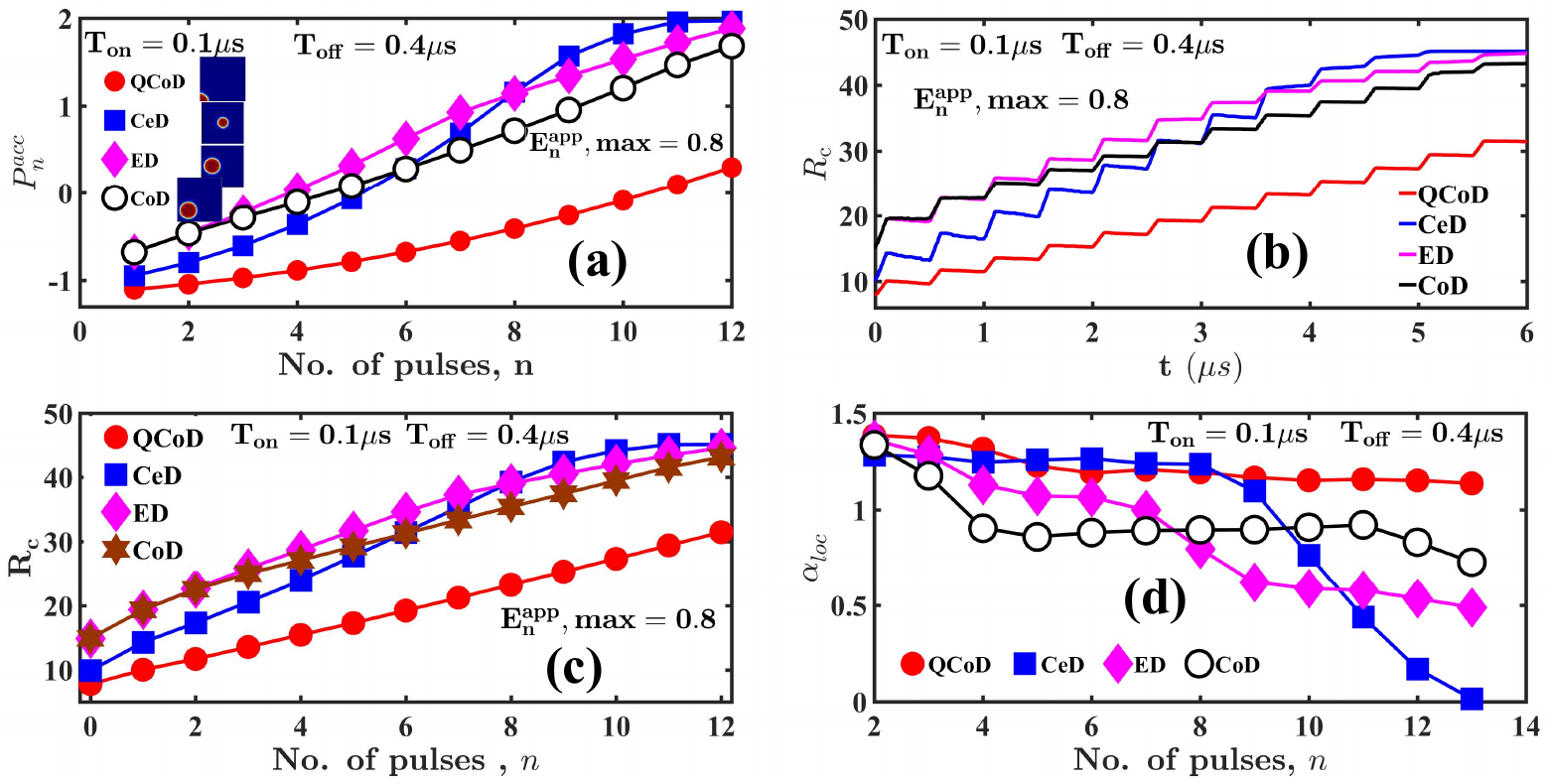}
	\vspace{-2.6cm}
	\caption{Comparative analysis of pulse-driven polarization accumulation kinetics for four FE systems with different initial domain configurations under a sequential voltage pulse train with $T^{}_{\mathrm{on}} = 0.1$ $\mu\mathrm{s}$, $T^{}_{\mathrm{off}} = 0.4$ $\mu\mathrm{s}$, and $E^{\rm app}_{n}, {\rm max} = 0.8$. (a) Evolution of the normalized accumulated polarization $P_n^{\mathrm{acc}}$ as a function of pulse number $n$. (b) Temporal evolution of the effective switched domain radius $R^{}_c$ during the pulse sequence, highlighting the stepwise domain expansion induced by successive voltage pulses. (c) Variation of the cumulative switched-domain radius $R^{}_c$ with $n$, demonstrating distinct accumulation kinetics arising from different initial domain configurations. (d) Evolution of the local growth exponent $\alpha_{\mathrm{loc}}$ as a function of $n$, revealing the transition between different domain-growth regimes and emphasizing the influence of initial domain geometry on pulse-driven switching dynamics.}
	\label{Figure3}
\end{figure*}

To quantitatively understand the effect of the initial pinned-domain configurations on polarization accumulation-mediated switching characteristics, we then investigate the evolution of the accumulated polarization $P^{acc}_{n}$, domain radius $R^{}_c$, and the corresponding $\alpha_{\rm local}$ for all the four configurations [as discussed Fig.~(\ref{Figure2})] in Fig.~(\ref{Figure3}). We have shown variation of $P^{\rm acc}_{n}$ as a function of $n$ for all the cases in Fig.~\ref{Figure3}(a). There is an increase in  $P^{\rm acc}_{n}$ in all the systems, indicating a finite increase in switched polarization with pulse-on, whereas only partial relaxation occurs during the pulse-off interval. Interestingly, $P^{\rm acc}_{n}$ varies linearly with $n$ for QCoD configurations, while for other cases its variation is parabolic (non-linear increment)  until domain walls do not hit the system boundary. Consequently, the system having CeD configuration exhibits the fastest polarization accumulation, followed by ED, CoD and QCoD. These observations also support the fact that the initial domain geometry primarily dictates pulse-driven switching by controlling the available domain-wall length and its interaction with the system boundaries.

The microscopic origin of this fascinating behaviour is further revealed in Fig.~\ref{Figure3}(b), where we analyze $R^{}_c$ variation as a function of $t$. The displacement of the domain wall towards the system boundary during each pulse exceeds its backward displacement, resulting in a net monotonous increase in $R^{}_c$(Type-III excitation dominates). To probe domain-growth kinetics in greater detail, we plot $R^{}_c$ as a function of $n$ in Fig.~\ref{Figure3}(c). As expected, the CeD configuration exhibits the most rapid increase in $R^{}_c$ due to the symmetrical positioning of the initially pinned domain, allowing nearly isotropic radial propagation during every pulse. On the contrary, the domain wall propagates asymmetrically in ED as an early interaction happens on one side of the domain wall with the sample boundary. The CoD configuration exhibits further reduction in the growth rate of $R^{}_c$ due to
constrained domain-wall motion, as it has two intersecting boundaries, whereas the QCoD configuration displays the slowest radial expansion due to the severe geometrical confinement of the initial domain. 
Fig.~\ref{Figure3}(d) shows the $\alpha^{}_{\rm local}$ variation as a function of $n$, providing an instantaneous measure of domain-growth kinetics. In the initial switching stage, both CeD and ED configurations show $\alpha^{}_{\rm local}>1$, indicating a super-linear growth regime where domain expansion accelerates with the pulse, which is mainly dictated by Type-I and Type-II excitation. While $\alpha^{}_{\rm local}$ decreases below $1$ at a later stage, reflecting a shift from accelerated to decelerating growth. It can be attributed to the depletion of negatively polarized regions and increased boundary interactions, reducing the dominance of Type-III excitation. Among all the systems, the CeD configuration exhibits the most pronounced acceleration, quickly transitioning to saturation due to unrestricted propagation.  As expected, the QCoD remains close to $\alpha^{}_{\rm local}\approx1$, indicating consistent growth kinetics due to geometrical confinement.

\begin{figure*}
	\includegraphics[scale=0.65]{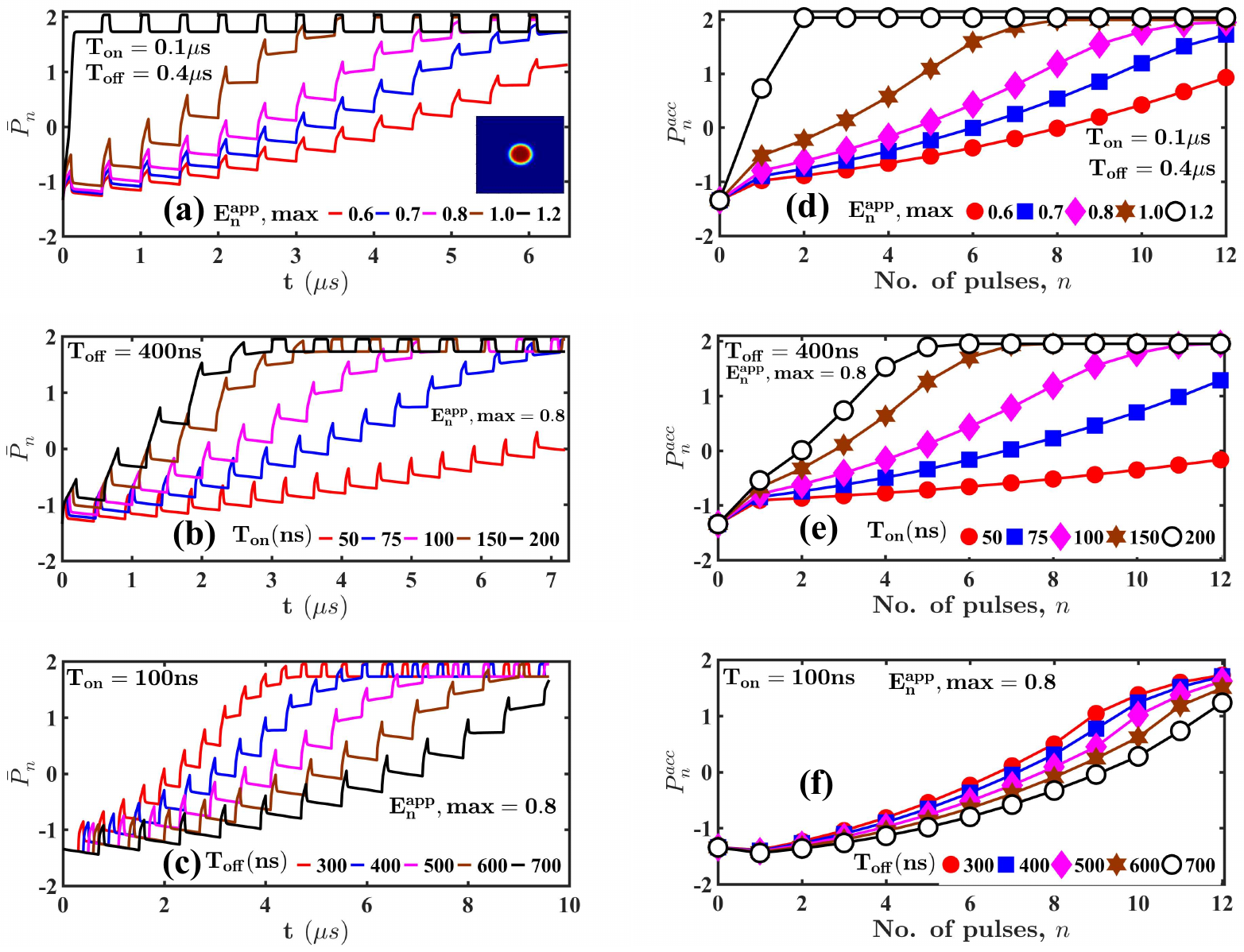}
	\caption{(a) Temporal evolution of the normalized polarization $\bar{P}^{}_n$ under sequential voltage pulses for different applied electric-field amplitudes $E^{\rm app}_{n}, {\rm max} = 0.6$, 0.7, 0.8, 1.0, and 1.2 with fixed $T^{}_{\mathrm{on}} = 0.1$ $\mu\mathrm{s}$ and $T^{}_{\mathrm{off}} = 0.4$ $\mu\mathrm{s}$ for the CeD configuration. (b) Influence of pulse-on time $T^{}_{\mathrm{on}}$ on the temporal evolution of $\bar{P}_n$ at fixed $E^{\rm app}_{n}, {\rm max}=0.8$ and $T^{}_{\mathrm{off}} = 400$ $\mathrm{ns}$. (c) Effect of pulse-off time $T_{\mathrm{off}}$ on the polarization dynamics for $T^{}_{\mathrm{on}} = 100$ $\mathrm{ns}$ and $E^{\rm app}_{n}, {\rm max} = 0.8$. (d–f) Corresponding accumulated polarization $P_n^{\mathrm{\rm acc}}$ as a function of pulse number $n$ varying (d) $E^{\rm app}_{n}, {\rm max}$ (e) $T^{}_{\mathrm{on}}$ and (f) $T^{}_{\mathrm{off}}$, respectively. It demonstrates that stronger electric fields and longer pulse-on durations accelerate polarization accumulation, whereas longer pulse-off durations suppress accumulation by promoting polarization relaxation between successive pulses.}
	\label{Figure4}
\end{figure*}

To understand the effect of pulse parameters variation on polarization accumulation, we analyze the variation of $\bar{P}^{}_{n}$ and $P^{\rm acc}_n$ for the CeD configuration in Fig.~(\ref{Figure4}). We have plotted $\bar{P}^{}_{n}$ vs. $t$ for five representative values of $E^{\rm app}_{n}, {\rm max}=0.6, 0.7, 0.8, 1.0$, and 1.2; and $T^{}_{\rm on}=0.1~\mu$s, $T^{}_{\rm off}=0.4~\mu$s in Fig.~\ref{Figure4}(a). For all field strengths,  $\bar{P}^{}_{n}$ exhibits the characteristic staircase-like evolution; there is also a net upward polarization jump with $E^{\rm app}_{n}, {\rm max}$. The polarization accumulates  slowly at lower field $E^{\rm app}_{n}, {\rm max}=0.6$, as only a small fraction of the negatively polarized domain undergoes irreversible switching during each pulse. The driving force responsible for domain-wall propagation becomes progressively stronger with an increase in $E^{\rm app}_{n}, {\rm max}$, leading to larger forward domain-wall displacements. As a consequence, Type-III excitation increasingly dominates over Type-III relaxation, resulting in progressively faster polarization accumulation at a later stage. Interestingly, there is an abrupt domain switching within the first few pulses with $E^{\rm app}_{n}, {\rm max}=1.2$, as it substantially diminishes the free-energy barrier, enabling rapid irreversible domain-wall propagation towards the system boundary.

 The influence of electric-field strength on polarization accumulation is further quantified in Fig.~\ref{Figure4}(d), where $P^{\rm acc}_{n}$ is plotted as a function of $n$. There is a clear transition from slow to rapid polarization accumulation with increasing $E^{\rm app}_{n}, {\rm max}$. It is because the increasing electric field enhances irreversible switching and weakens the relative influence of spontaneous relaxation. These observations indicate that the applied electric field primarily controls the efficiency of Type-III excitation, which ultimately governs the polarization accumulation kinetics. In Fig.~\ref{Figure4}(b) and Fig.~\ref{Figure4}(e), we investigate the effect of $T^{}_{\rm on}$ variation on polarization accumulation characteristics with $E^{\rm app}_{n}, {\rm max}=0.8$ and $T^{}_{\rm off}=400$ ns. There is a monotonous increment in polarization with $T^{}_{\rm on}$. Besides, the duration of field-driven Type-III excitation becomes progressively dominant with $T^{}_{\rm on}$, allowing the domain wall to advance farther during the pulse-on interval. It results in net polarization increments after every pulse and hence accelerates polarization accumulation. For the largest $T^{}_{\rm on}=200$ ns, nearly complete switching is achieved with only a few pulses, suggesting that sufficiently long excitation duration can suppress spontaneous relaxation, driving the system toward saturation. The influence of the pulse-off time $T^{}_{\rm off}$ is analyzed in Fig.~\ref{Figure4}(c) and Fig.~\ref{Figure4}(f) with $T^{}_{\rm off}$ varied between 300 and 700 ns while keeping $T^{}_{\rm on}=100$ ns and $E^{\rm app}_{n}, {\rm max}=0.8$. As expected, increasing $T^{}_{\rm off}$ has a diminishing effect on polarization accumulation. It is because the absence of the field during the pulse-off interval allows the system to evolve solely under the influence of internal thermodynamic driving forces. Consequently, spontaneous relaxation becomes increasingly important and hence enhancement in the backward displacement of the domain wall with increasing $T^{}_{\rm off}$. These observations clearly show that while $E^{\rm app}_n$ and $T^{}_{\rm on}$ primarily control the pulse-driven excitation, $T^{}_{\rm off}$ governs the extent of spontaneous relaxation.

\begin{figure*}
	\includegraphics[scale=0.65]{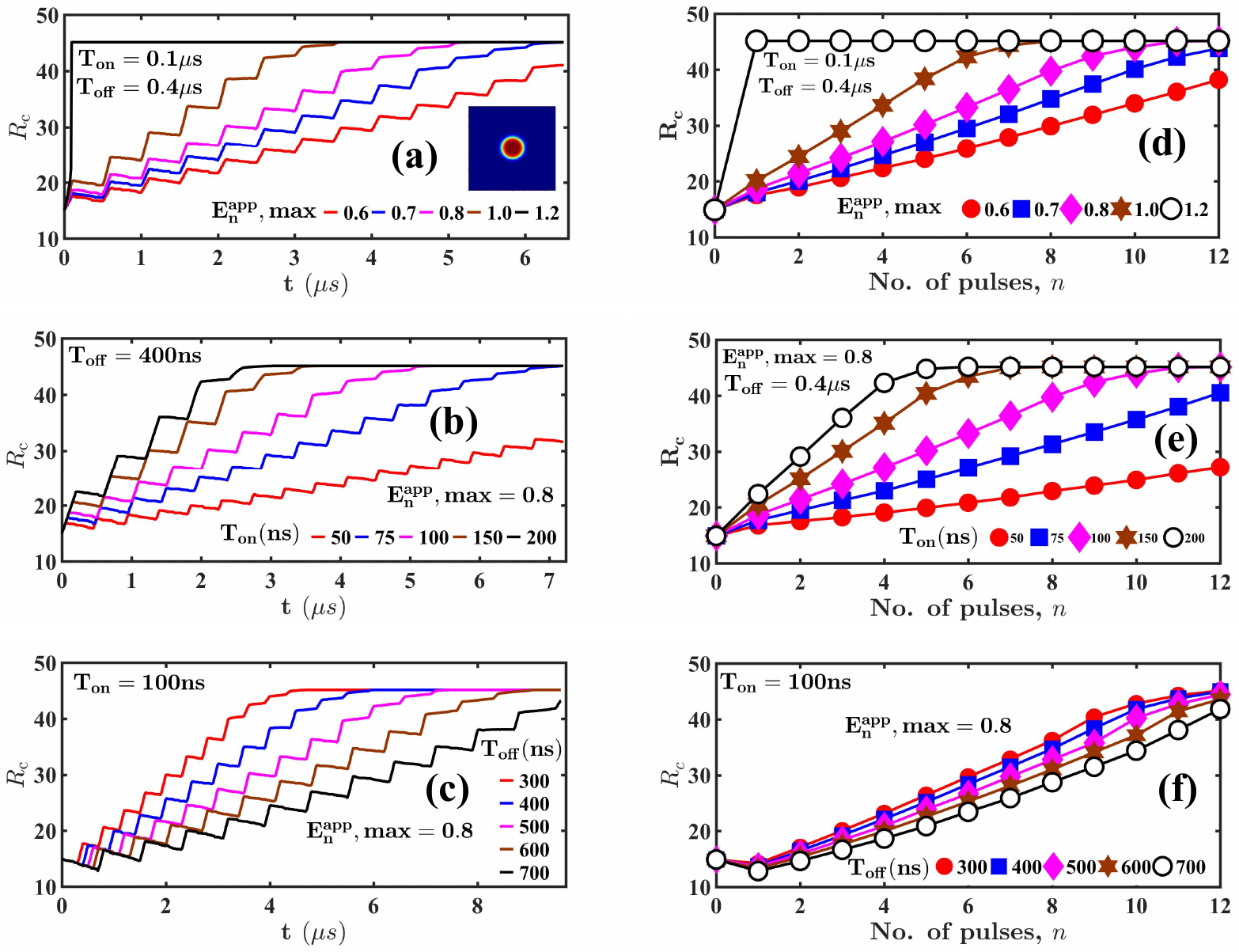}
	\caption{Evolution of the effective switched-domain radius $R^{}_c$ during pulse-driven accumulative polarization switching under different pulse parameters for the CeD configuration. (a–c) Temporal evolution of $R^{}_c$ by varying (a) applied electric-field amplitude $E^{\rm app}_{n}, {\rm max}$, (b) pulse-on time $T^{}_{\mathrm{on}}$, and (c) pulse-off time $T^{}_{\mathrm{off}}$. The inset in (a) shows the CeD FE system used for this study. (d–f) Corresponding evolution of $R^{}_c$ as a function of pulse number $n$, illustrating the cumulative domain-growth kinetics for different (d) $E^{\rm app}_{n}, {\rm max}$, (e) $T^{}_{\mathrm{on}}$, and (f) $T^{}_{\mathrm{off}}$. Increasing the electric-field amplitude or pulse-on time significantly accelerates domain expansion and promotes faster accumulation, whereas increasing the pulse-off time suppresses the growth rate by allowing enhanced polarization relaxation between successive pulses.}
	\label{Figure5}
\end{figure*}

To further unravel the microscopic origin of the accumulation-mediated polarization switching, we study  the evolution of $R^{}_c$ for the CeD configuration for various relevant parameters in Fig.~(\ref{Figure5}). The first column presents the evolution of $R^{}_c$ as a function of $t$, whereas the second column shows the corresponding variation with $n$. All the parameter values are the same as those of Fig.~(\ref{Figure4}). We investigate the influence of $E^{\rm app}_{n}, {\rm max}$ on $R^{}_c$ in Fig.~\ref{Figure5}(a), an important measure of 
domain-growth kinetics. There exhibits a staircase-like evolution of  $R^{}_c$ for all $E^{\rm app}_{n}, {\rm max}$, indicating continuous domain expansion dictated by the applied field pulse followed by slight contraction during the pulse-off interval. The driving force is relatively weaker at lower field amplitude $E^{\rm app}_{n}, {\rm max}=0.6$, resulting in limited outward domain-propagation during the pulse-on interval. Consequently, only a slight increase in $R^{}_c$ is retained as spontaneous relaxation partially compensates the preceding domain growth during the pulse-off period. Since the forward Type-III excitation increasingly dominates over the spontaneous Type-III relaxation for larger $E^{\rm app}_{n}, {\rm max}=0.8$, $R^{}_c$ grows much more rapidly, eventually reaching complete saturation within only a few pulses at the highest applied field $E^{\rm app}_{n}, {\rm max}=1.2$.

The effect of $T^{}_{\rm on}$ variation on $R^{}_c$ is presented in Fig.~\ref{Figure5}(b) and Fig.~\ref{Figure5}(e). There is a monotonous enhancement in the domain-growth rate with $T^{}_{\rm on}$ as the domain wall experiences the driving force for a longer period. Specifically for small $T^{}_{\rm on}$, the net increase in the $R^{}_c$ after each pulse remains relatively small because the field is removed eventually much before the domain wall can propagate significantly, allowing spontaneous relaxation  to compensate a considerable fraction of the preceding growth during the subsequent pulse-off period. In contrast, longer pulse widths (larger $T^{}_{\rm on}$) provide sufficient time for sustained domain-wall propagation, enabling larger radial expansion, resulting in cumulative domain growth. Fig.~\ref{Figure5}(c) and Fig.~\ref{Figure5}(f) show the influence of $T^{}_{\rm off}$ on the domain-growth kinetics. Interestingly, $R^{}_c$ gets suppressed with $T^{}_{\rm off}$ because during the pulse-off period, the absence of the external electric field allows the system to evolve under the influence of internal thermodynamic driving forces, resulting in the dominance of spontaneous relaxation of the domain wall. As a consequence, the domain wall has more time to retract toward its equilibrium position with larger $T^{}_{\rm off}$, producing a significant reduction in $R^{}_c$. 

\begin{figure*}
	\includegraphics[scale=0.65]{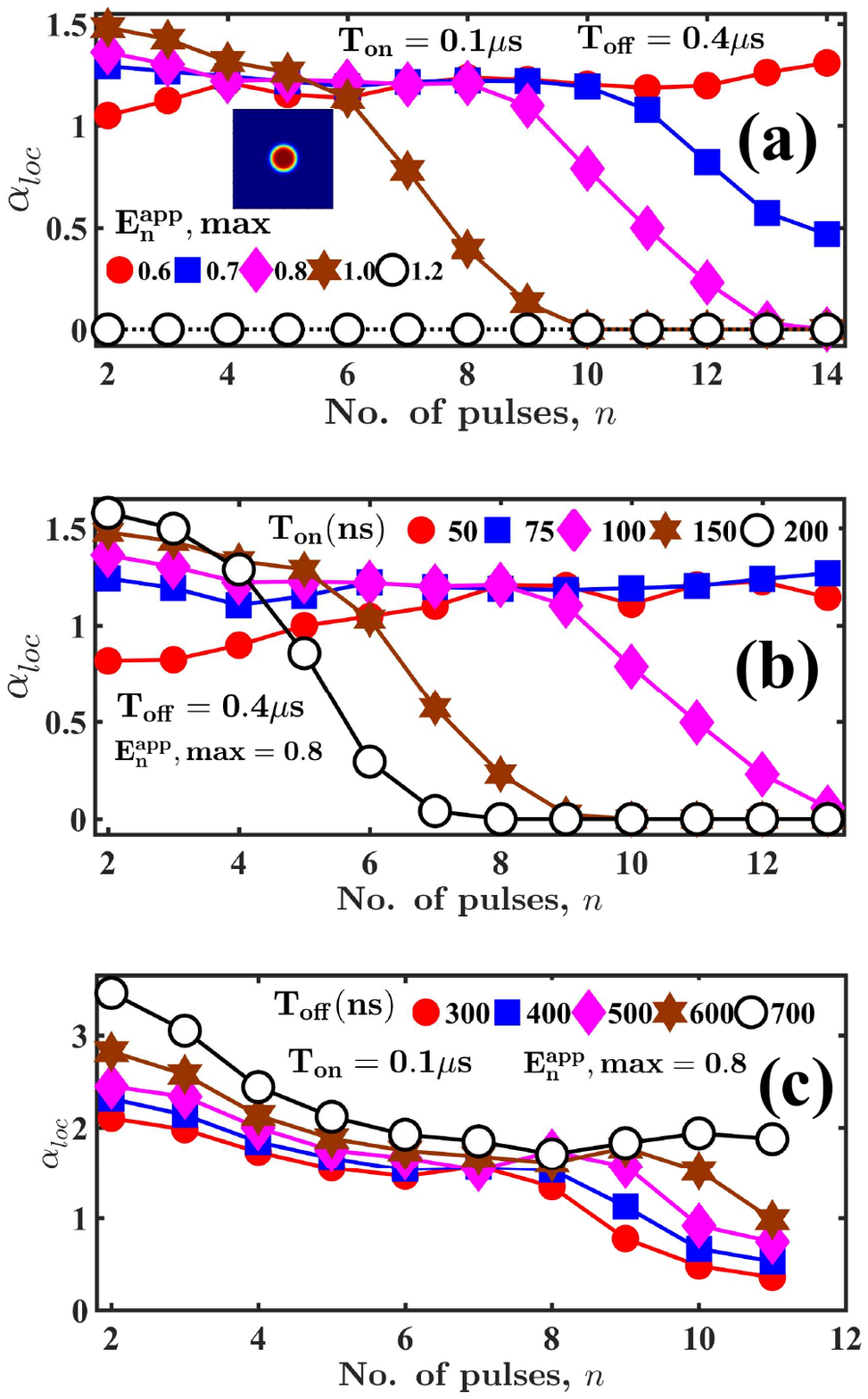}
	\caption{Evolution of the local growth exponent $\alpha^{}_{\mathrm{loc}}$ during pulse-driven accumulative polarization switching under different pulse conditions. (a) Variation of $\alpha^{}_{\mathrm{loc}}$ with pulse number $n$ for different $E^{\rm app}_{n}, {\rm max}$ at fixed $T^{}_{\mathrm{on}}=0.1$ $\mu\mathrm{s}$ and $T^{}_{\mathrm{off}}=0.4$ $\mu\mathrm{s}$. (b) Dependence of $\alpha^{}_{\mathrm{loc}}$ on $n$ for different $T^{}_{\mathrm{on}}$ with fixed $E^{\rm app}_{n}, {\rm max}=0.8$  and $T^{}_{\mathrm{off}}=0.4$ $\mu\mathrm{s}$. (c) Evolution of $\alpha^{}_{\mathrm{loc}}$ for various $T^{}_{\mathrm{off}}$ at fixed $E^{\rm app}_{n}, {\rm max}=0.8$  and $T^{}_{\mathrm{on}}=0.1$ $\mu\mathrm{s}$. It unravels the evolution of the instantaneous domain-growth kinetics, revealing transitions between accelerated growth and relaxation-dominated regimes. Stronger $E^{\rm app}_{n}, {\rm max}$ and longer $T^{}_{\mathrm{on}}$ promote an earlier crossover to the relaxation-controlled regime, whereas increasing the $T^{}_{\mathrm{off}}$ delays polarization accumulation.}
	\label{Figure6}
\end{figure*}

To obtain a deeper understanding of the domain-growth dynamics during  polarization accumulation, Fig.~(\ref{Figure6}) presents the evolution of the local kinetic exponent $\alpha^{}_{\rm local}$ as a function of various parameters of interest. It could provide an instantaneous measure of the domain-growth kinetics and therefore enable the identification of different kinetic regimes during polarization accumulation-driven switching. Fig.~\ref{Figure6}(a)–Fig.~\ref{Figure6}(c) illustrate the influence of $E^{\rm app}_{n}, {\rm max}$, $T^{}_{\rm on}$, and $T^{}_{\rm off}$ on the evolution of $\alpha^{}_{\rm local}$. It is quite clear that $\alpha^{}_{\rm local}$ increases rapidly during the initial switching stage for large $E^{\rm app}_{n}, {\rm max}$. Such super-linear growth indicates that the switching kinetics get accelerated with successive pulses, suggesting  that each pulse becomes progressively more effective in driving domain-wall propagation than the preceding one. It can also be understood as follows: successive voltage pulses progressively destabilize the surrounding negative polarization through Type-I excitation while simultaneously instigating the already switched positive domain to go through Type-II excitation. As a result, the energy barrier for domain-wall propagation is continuously reduced, allowing the irreversible Type-III excitation to become dominant, hence dictating the polarization-accumulation dynamics.   

 Increasing $E^{\rm app}_{n}$ significantly enhances $\alpha^{}_{\rm local}$, indicating a stronger acceleration of the switching kinetics. At higher electric fields, the thermodynamic driving force promotes faster domain-wall propagation, allowing successive pulses to produce increasingly larger domain wall displacements. Similarly, increasing $T^{}_{\rm on}$ extends the duration of the electric field acting on the domain-wall, thereby sustaining forward propagation for longer periods and hence maximizing $\alpha^{}_{\rm local}$. In contrast, there is suppression of the acceleration regime with $T^{}_{\rm off}$ because the domain wall undergoes greater spontaneous relaxation. Consequently, the enhancement of domain-wall mobility produced during pulse-on is only partially retained, resulting in comparatively smaller values of $\alpha^{}_{\rm local}$. Surprisingly, $\alpha^{}_{\rm local}$ gradually decreases, indicating a transition to a nearly linear domain-growth regime at a later stage. It is because the domain wall advances by approximately the same distance during successive pulse cycles, reaching a dynamic balance. Although Type-III excitation continues to dominate over relaxation, the efficiency of successive pulses no longer increases significantly because the switched domain has already expanded over a considerable fraction of the system. The transition from $\alpha^{}_{\rm local}>1$ to $\alpha^{}_{\rm local}<1$ therefore signifies a fundamental evolution of the switching mechanism. During the early pulses, the domain-wall propagation is excitation-dominated. During the intermediate stage, the system exhibits near-linear propagation, corresponding to a dynamic balance between the enhancement of excitation and the onset of geometric constraints. Finally, the system enters a deceleration or saturation regime at much later stages, where finite-size effects, depletion of the remaining switchable polarization, and spontaneous relaxation collectively suppress further acceleration of the domain wall.

\begin{figure*}
	\includegraphics[scale=0.65]{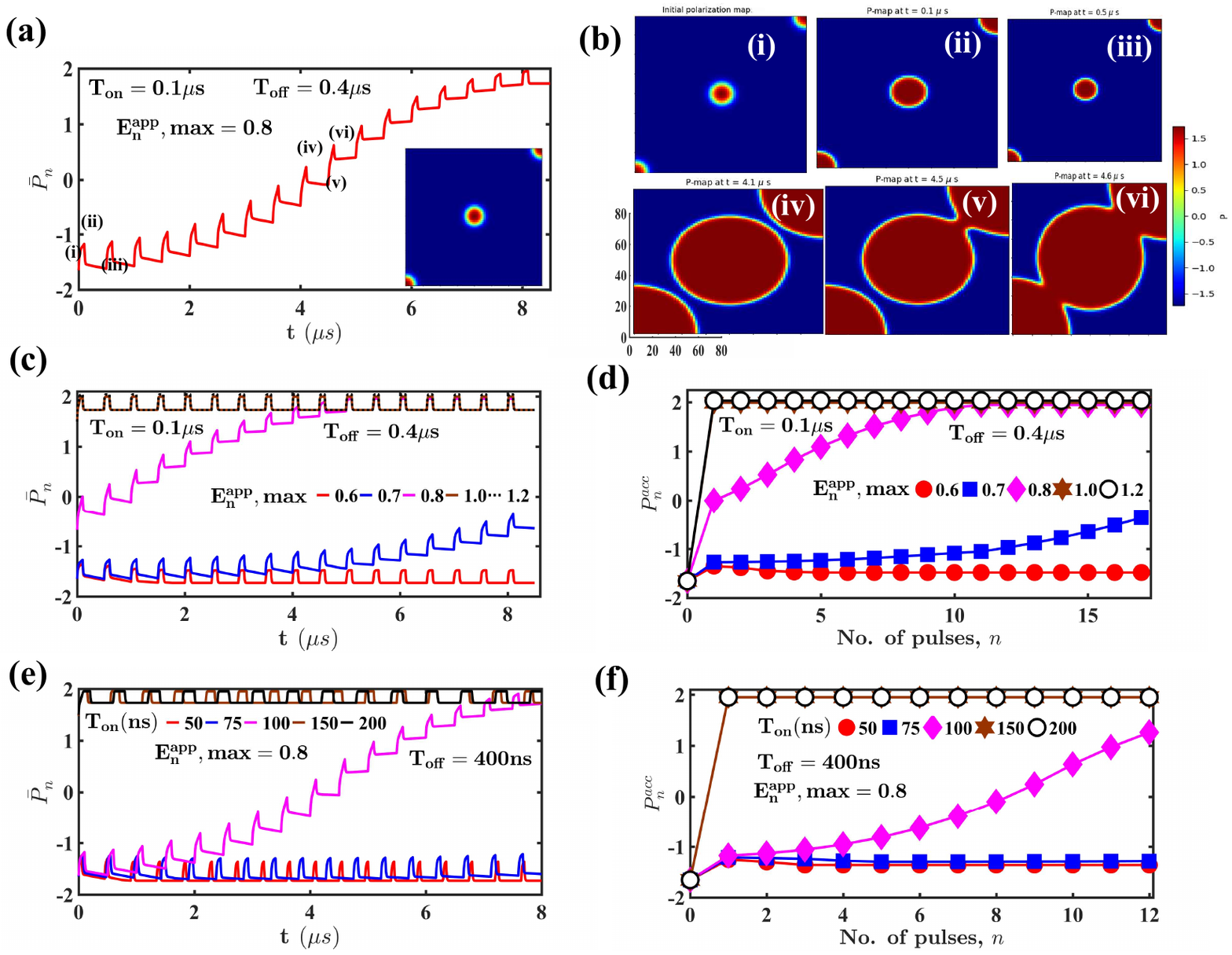}
	\caption{(a) Temporal evolution of $\bar{P}_n$ for $T^{}_{\mathrm{on}}=0.1$ $\mu\mathrm{s}$, $T^{}_{\mathrm{off}}=0.4$ $\mu\mathrm{s}$, and $E^{\rm app}_{n}, {\rm max}=0.8$ for Diagonal Corner–Centre Configuration (DCC) configuration (b) Corresponding polarization maps at the instants indicated in (a), illustrating the nucleation, pulse-induced domain expansion, partial relaxation during the pulse-off interval, renewed domain growth under subsequent pulses. (c, e) Comparison of $\bar{P}^{}_{n}$ vs. $t$ for different $E^{\rm app}_{n}, {\rm max}$ and $T^{}_{\mathrm{on}}$, respectively. (d, f) Corresponding accumulated polarization $P^{\rm acc}_{n}$ as a function of pulse number $n$. It clearly demonstrates that stronger $E^{\rm app}_{n}, {\rm max}$ and longer $T^{}_{\mathrm{on}}$  durations accelerate the transition from incomplete to complete accumulative switching by enhancing domain-growth kinetics and suppressing the influence of relaxation.} 
	\label{Figure7}
\end{figure*}

To investigate the influence of multiple initially pinned domains on pulse-driven accumulative switching, we now investigate the polarization evolution of the Diagonal Corner–Centre Configuration (DCC): two domains are located at diagonally opposite corners in addition to centrally positioned circular domains in the FE system in Fig.~(\ref{Figure7}). Such a simple configuration surprisingly introduces the simultaneous propagation of two independent switching fronts (cornered and centrally-positioned domains), opening an elegant pathway to understand the influence of domain-domain interaction on polarization accumulation. Fig.~\ref{Figure7}(a) shows the time evolution of $\bar{P}^{}_n$, which exhibits a staircase-like evolution similar to the single-domain configurations.  However, the polarization increases more rapidly as compared with a single-domain system because the centrally pinned domain expands almost isotropically during the initial pulses, whereas the corner domain also propagates toward the interior due to the restriction imposed by the boundary on its outward motion. As the pulse sequence proceeds, the cumulative effect of repeated excitation progressively increases the switched area until the two expanding domains approach one another (spontaneous excitation), resulting in a rapid polarization accumulation. 

The polarization morphologies are also presented at typical time instants in Fig.~\ref{Figure7}(b), to provide direct insight into the microscopic switching mechanism. It is quite clear that Type-I excitation dictates in the negatively polarized matrix initially, resulting in a positive polarization increment without reversing orientation while the positively polarized matrix undergoes Type-II excitation. On the other hand, during every pulse-off interval, Type-I and Type-II relaxation restore the polarization magnitudes toward their remanent states, while Type-III relaxation induces a slight backward motion of both domain walls. Since the forward propagation during the pulse-on interval exceeds the backward relaxation, both domains continue to expand with increasing pulse number. Consequently, the central domain expands almost uniformly in all directions, whereas the corner domain exhibits asymmetric growth in such an exotic system. The interaction between the two expanding domains becomes increasingly important as switching progresses. In due time, the repeated pulse excitation progressively destabilizes the negatively polarized bridge separating the two domains through cumulative Type-I excitation, while repeated Type-II excitation further stabilizes the already switched regions. Consequently, the driving force for Type-III excitation increases, resulting in accelerated propagation of both switching fronts toward each other, ultimately resulting in coalescence of the domains. 
The influence of various parameters on $\bar{P}^{}_{n}$ evolution is then shown in Fig.~\ref{Figure7}(c) and Fig.~\ref{Figure7}(e). Increasing either the $E^{\rm app}_{n}, {\rm max}$ or $T^{}_{\rm on}$ significantly enhances the upward polarization increment by strengthening the driving force and increasing Type-III excitation-induced domain-wall propagation. 
The corresponding accumulated polarization $P^{\rm acc}_{n}$ as a function of pulse number $n$ is shown in Fig.~\ref{Figure7}(d) and Fig.~\ref{Figure7}(f). There is always a monotonic enhancement in $P^{\rm acc}_{n}$, confirming that the pulse-driven excitation consistently dominates over spontaneous relaxation with $E^{\rm app}_{n}, {\rm max}$. Conversely, increasing $T^{}_{\rm off}$ progressively delays the accumulation process because spontaneous Type-III relaxation becomes increasingly significant. Although reversible Type-I and Type-II excitation and relaxation continuously occur throughout the switching process, the accumulated polarization is primarily determined by the competition between forward and backward Type-III domain-wall motion.

\begin{figure*}
	\includegraphics[scale=0.70]{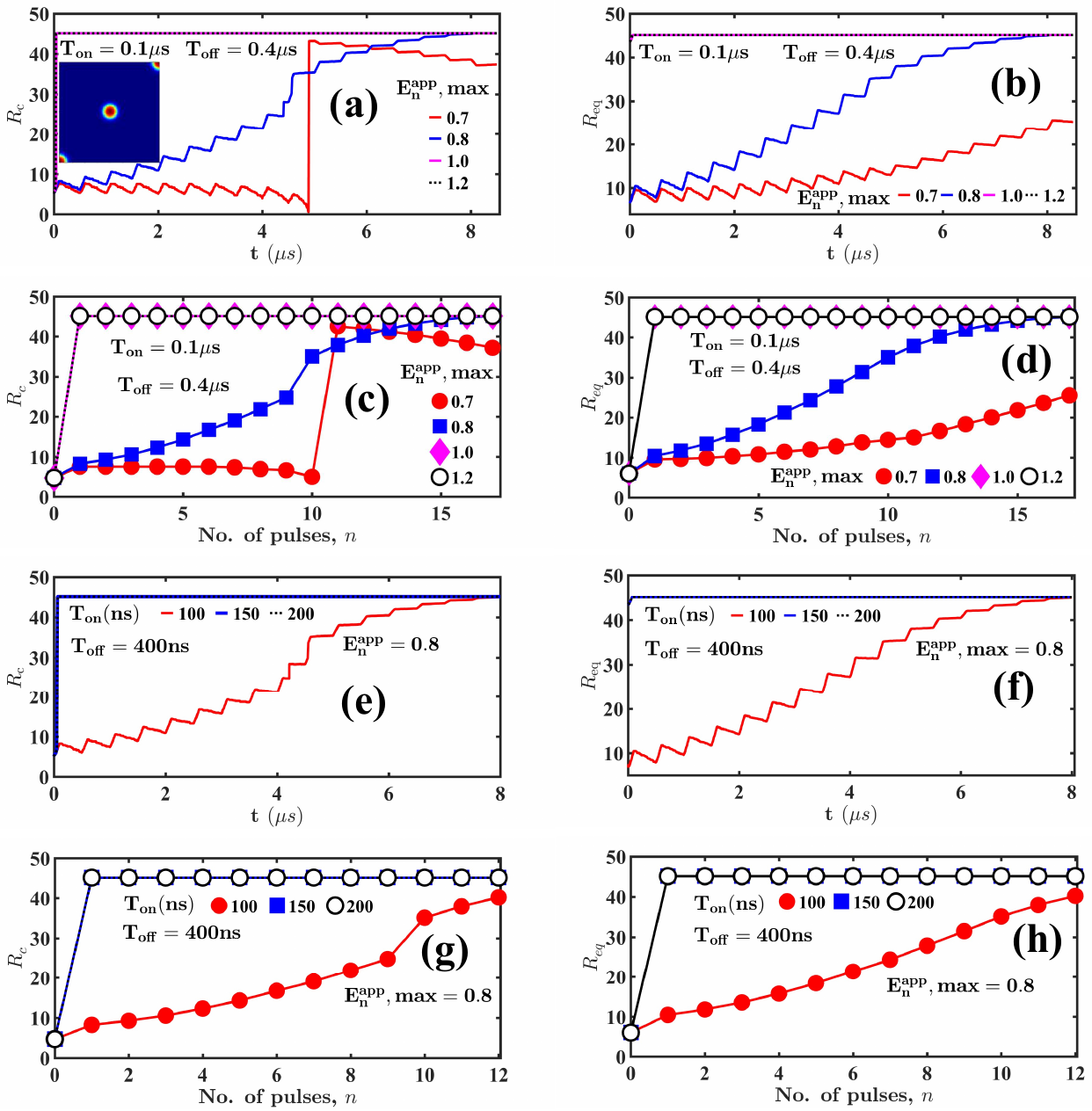}
	\caption{(a, b) Temporal evolution of the central-domain radius $R^{}_c$ and the equivalent switched-domain radius $R^{}_{\rm eq}$, respectively, for different $E^{\rm app}_{n}, {\rm max}$ at fixed $T^{}_{\rm on}=0.1$ $\mu\mathrm{s}$ and $T^{}_{\rm off}=0.4$ $\mu\mathrm{s}$. (c, d) Corresponding variations of $R^{}_c$ and $R^{}_{\mathrm{eq}}$ with pulse number $n$, illustrating the scaling behaviour of individual-domain and collective multi-domain growth. (e,f) Temporal evolution of $R^{}_c$ and $R^{}_{\rm eq}$, respectively, for different $T^{}_{\rm on}$ with fixed $E^{\rm app}_{n}, {\rm max}=0.8$ and $T^{}_{\rm off}=0.4$ $\mu\mathrm{s}$. (g,h) Corresponding dependence of $R^{}_c$ and $R^{}_{\rm eq}$ on $n$ for different $T^{}_{\rm on}$. The staircase-like evolution of both radii reflects the competition between field-driven domain-wall propagation during the pulse-on interval and partial relaxation during the pulse-off interval. Stronger electric fields and longer pulse-on durations accelerate multi-domain propagation and enhance cooperative polarization accumulation.} 
	\label{Figure8}
\end{figure*}

To further elucidate the influence of multiple initially pinned domains on the switching kinetics, Fig.~(\ref{Figure8}) presents the evolution of the domain radius for the DCC configurations. Here we calculate the domain radius in two ways: (i) $R^{}_c$ represents the radius of only the centrally pinned circular domain and therefore quantifies its intrinsic propagation kinetics. (ii) The second measure, $R^{}_{\rm eq}$, is calculated by considering all three domains and thus characterizes the overall evolution of the multi-domain system. This distinction enables us to investigate the individual growth of the central domain as well as collective switching behaviour arising from simultaneous propagation of all three domains separately. In Fig.~\ref{Figure8}(a) and Fig.~\ref{Figure8}(b), we show the temporal evolution of $R^{}_c$ and $R^{}_{\rm eq}$, respectively, for various values of $E^{\rm app}_{n}, {\rm max}$. Both $R^{}_c$ and $R^{}_{\rm eq}$ exhibit the characteristic staircase-like evolution, irrespective of pulse parameters. However, the behaviour of $R^{}_c$ differs significantly from that of $R^{}_{\rm eq}$, explained below. Interestingly, the central domain (quantified by $R^{}_c$) initially exhibits only modest growth as Type-III excitation is nearly balanced by spontaneous Type-III relaxation at small $E^{\rm app}_{n}, {\rm max}$. In contrast, $R^{}_{\rm eq}$ increases continuously because it includes the simultaneous expansion of the two corner domains as well as the central domain, which collectively contribute to the total switched area. As the electric field increases, the domain-wall mobility of all three domains increases substantially, resulting in rapid outward propagation and significantly faster growth of both $R^{}_c$ and $R^{}_{\rm eq}$. At sufficiently high electric fields, complete switching is achieved within only one or two pulses because the free-energy barrier for polarization reversal is effectively eliminated.

The dependence of $R^{}_c$ and $R^{}_{\rm eq}$ on pulse number $n$ is presented in Fig.~\ref{Figure8}(c) and Fig.~\ref{Figure8}(d). The evolution of $R^{}_c$ clearly indicates that the growth of the central domain is highly non-linear. Initially, the domain expands gradually, primarily driven by the motion of its circular domain wall. However, after several pulse cycles, the growth rate accelerates significantly as corner domains tend to coalesce with the central domain, effectively reducing the negatively polarized region. In such a case, repeated Type-I and Type II-excitation destabilizes and stabilizes the switched domains, respectively, concurrently. Consequently, the driving force for irreversible Type-III excitation comes into play, resulting in rapid outward propagation of the central domain until it merges seamlessly with the corner domains. In Fig.~\ref{Figure8}(e) and Fig.~\ref{Figure8}(f), we analyzed the effect of pulse-on duration $T^{}_{\rm on}$ on $R^{}_c$ and $R^{}_{\rm eq}$ as a function of $t$. There is a monotonous increase in domain size with $T^{}_{\rm on}$ due to enhancement in domain-wall propagation. In addition, large $T^{}_{\rm on}$ creates larger displacements, accelerating the expansion of domains. This is particularly reflected in $R^{}_c$ variation, where growth speeds up due to earlier interactions and coalescence. The increase in $R^{}_{eq}$ is even more significant, as the retained switched area promotes quicker accumulation of multi-domain polarization. Similar observations can be drawn from Fig.~\ref{Figure8}(g) and Fig.~\ref{Figure8}(h), where we show the variation of $R^{}_c$ and $R^{}_{\rm eq}$ with $n$.

\begin{figure*}
	\includegraphics[scale=0.65]{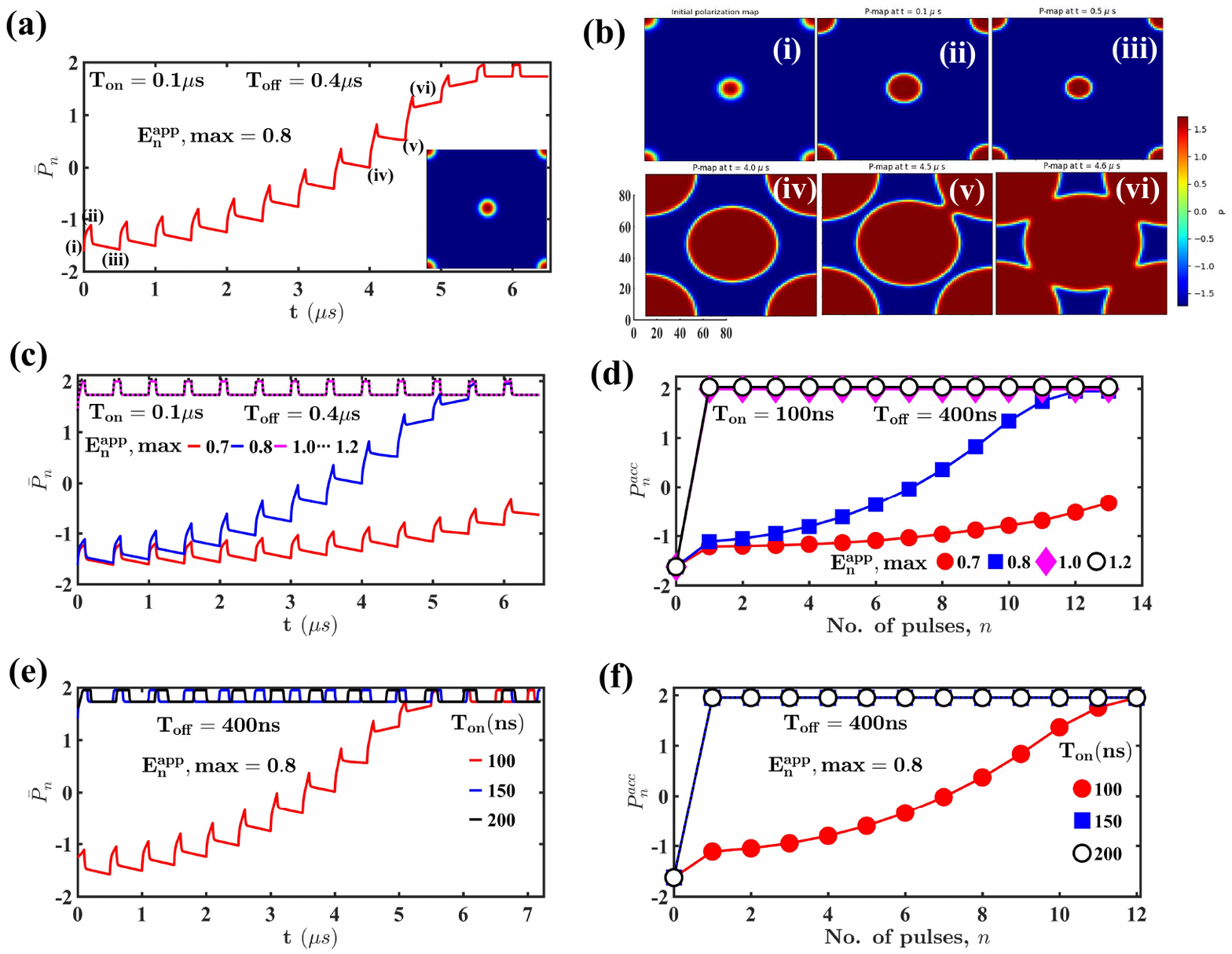}
	\caption{(a) Temporal evolution of $\bar{P}^{}_n$ under sequential voltage pulses with $T_{\rm on}=0.1$ $\mu\mathrm{s}$, $T^{}_{\rm off}=0.4$ $\mu\mathrm{s}$, and $E^{\rm app}_{n}, {\rm max}=0.8$ for Corner–Centre Configuration (CCC); inset shows the initial domain configuration. (b) Corresponding polarization maps at the instants marked in (a). (c,e)  $\bar{P}^{}_n$ vs. $t$ by $E^{\rm app}_{n}, {\rm max}$ and $T^{}_{\mathrm{on}}$, respectively. (d,f) Corresponding accumulated polarization $P_n^{\rm acc}$ as a function of pulse number $n$, demonstrating that stronger electric fields and longer pulse-on durations substantially accelerate polarization accumulation. Compared with the DCC configuration, the smaller initial domain separation enhances domain interactions, accelerating cumulative switching and reducing the number of pulses required for complete polarization reversal.} 
	\label{Figure9}
\end{figure*}

To further investigate the role of initial domain arrangement on pulse-driven accumulative switching, we study the polarization evolution for the Corner-Center Configuration (CCC) in Fig.~(\ref{Figure9}). In such an arrangement, we cleverly place one positively polarized circular domain at the centre of the FE system while the remaining domains are pinned at the four corners. It is natural to expect completely different polarization accumulation characteristics as compared with all the configurations considered before due to the reduced separation between the initial domains, leading to stronger domain-domain interaction and   affecting the switching process significantly. Fig.~\ref{Figure9}(a) shows the evolution of $\bar{P}^{}_{n}$ as a function of $t$. As expected, the polarization exhibits a staircase-like profile consisting of rapid increases during the pulse-on interval followed by partial relaxation during the pulse-off. However, compared with the DCC configuration, the polarization increases more rapidly because the neighbouring corner domains interact with the centrally positioned domain at an earlier stage of switching. Consequently, a larger fraction of the system undergoes irreversible polarization reversal (Type III-excitation) within the first few pulse cycles. The corresponding polarization morphologies are shown in Fig.~\ref{Figure9}(b), providing direct insight into the microscopic switching mechanism. During the initial stage, All the pinned domains evolve independently. Type-I and Type II-excitation/relaxation go on concurrently in the negative and positive-polarized matrix, respectively. On the other hand, Type-III excitation drives outward propagation of the domain walls. Consequently, the central circular domain expands almost isotropically, whereas the corner domains propagate preferentially toward the interior because the sample boundaries restrict outward motion. Since the forward propagation during the pulse-on interval consistently exceeds the backward relaxation, all domains continue to grow with time. This cooperative propagation instigates domain coalescence comparatively at a very early stage, leading to the formation of a larger connected positively polarized region. Once coalescence occurs, the internal interfaces between the individual domains disappear, and the subsequent switching is governed primarily by the outward propagation of the common outer domain boundary.

The influence of various parameters on the $\bar{P}^{}_{n}$ is examined in Fig.~\ref{Figure9}(c) and Fig.~\ref{Figure9}(e). The domain-wall propagation gets elevated with $E^{\rm app}_{n}, {\rm max}$ and $T^{}_{\rm on}$ due to an enhancement in driving force, which ultimately results in an increase in  $\bar{P}^{}_{n}$. On the other hand, a reduction in $\bar{P}^{}_{n}$ with $T^{}_{\rm off}$ is observed due to the domination of spontaneous relaxation. Although these trends are consistent with those observed in the previous configurations, the CCC system exhibits a stronger overall response because the more closely spaced domains interact strongly under repeated pulse excitation. Fig.~\ref{Figure9}(d) and Fig.~\ref{Figure9}(f) show $P^{\rm acc}_{n}$ as a function of $n$.  $P^{\rm acc}_{n}$ increases monotonically with $n$, confirming that irreversible switching progressively dominates over spontaneous relaxation. Larger $E^{\rm app}_{n}, {\rm max}$ and $T^{}_{\rm on}$ substantially reduce the number of pulses required to achieve complete switching by enhancing the efficiency of Type-III excitation and accelerating cooperative domain-wall propagation. Nevertheless, owing to the strong interaction between the neighbouring domains, the CCC configuration maintains relatively efficient and accelerated polarization accumulation even under less favourable pulse conditions.

\begin{figure*}
	\includegraphics[scale=0.70]{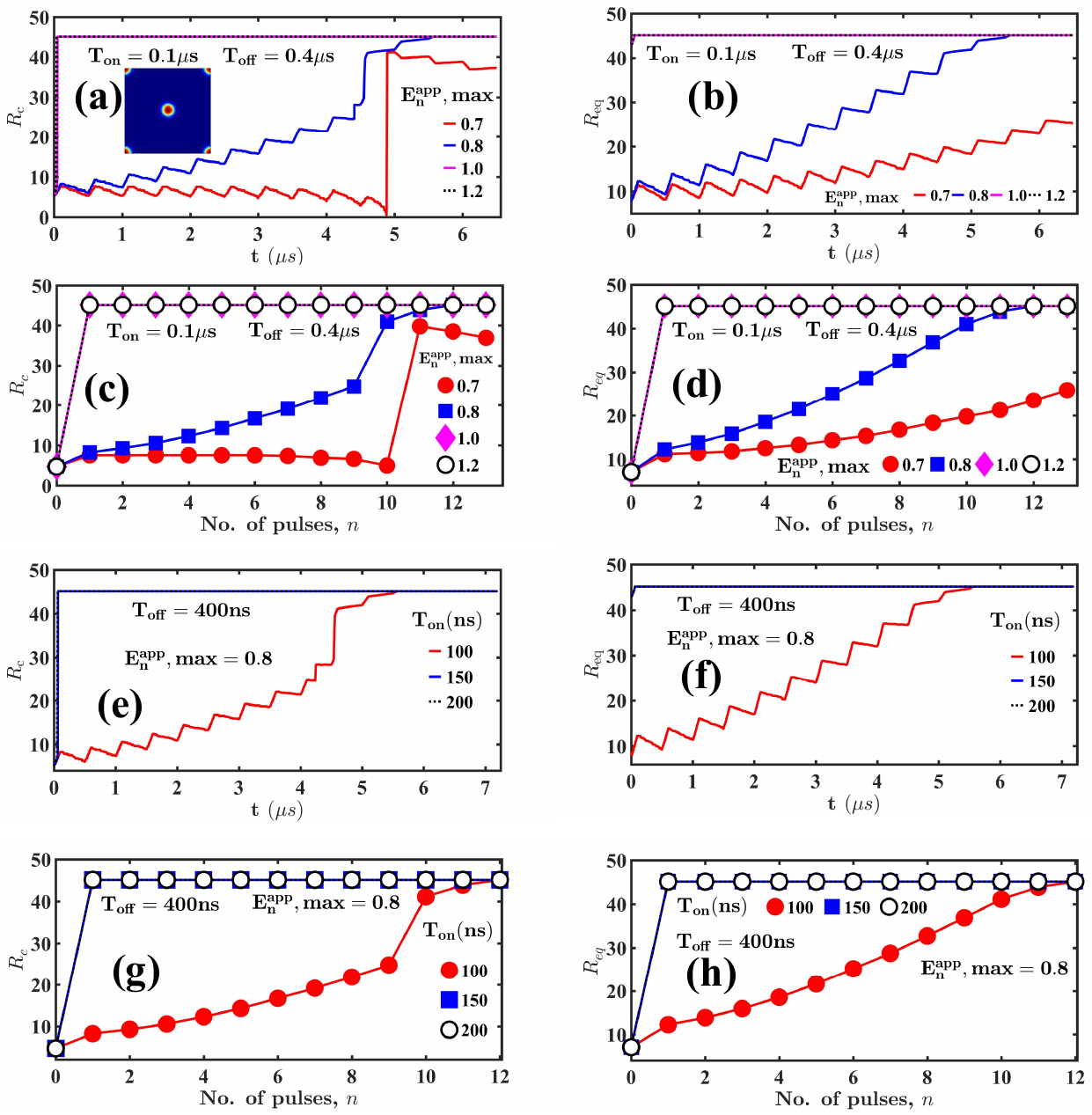}
	\caption{(a,b) Temporal evolution of $R^{}_c$ and $R^{}_{\rm eq}$, respectively, for different $E^{\rm app}_{n}, {\rm max}$ with fixed $T^{}_{\rm on}=0.1$ $\mu\mathrm{s}$ and $T^{}_{\rm off}=0.4$ $\mu\mathrm{s}$ for the CCC configuration. (c,d) Corresponding dependence of $R^{}_c$ and $R^{}_{\mathrm{eq}}$ on pulse number $n$. (e,f) Temporal evolution of $R^{}_c$ and $R_{\rm eq}$, respectively, for different $T^{}_{\rm on}$ at fixed $E^{\rm app}_{n}, {\rm max}=0.8$ and $T_{\rm off}=400$ $\mathrm{ns}$. (g,h) Corresponding variation of $R^{}_c$ and $R_{\rm eq}$ with $n$ for different $T_{\rm on}$. The staircase-like evolution in both radii is evident. Increasing the electric-field amplitude or pulse-on duration enhances irreversible domain-wall motion, accelerates cooperative multi-domain growth, and reduces the number of pulses required to achieve complete polarization switching.} 
	\label{Figure10}
\end{figure*}

To further quantify the multi-domain switching kinetics in the CCC configurations, Fig.~(\ref{Figure10}) presents the evolution of the domain radius as a function of both $t$ and $n$ under different pulse conditions. Fig.~\ref{Figure10}(a) and Fig.~\ref{Figure10}(b) show the temporal evolution of $R^{}_c$ and $R^{}_{\rm eq}$, respectively. Both radii exhibit the characteristic staircase-like evolution similar to that of other configurations discussed before. However, the growth of $R^{}_c$ is strongly influenced by the neighbouring corner domains. Initially, the central domain propagates almost isotropically. As switching progresses, the expanding corner domains approach the central domain much earlier owing to their shorter initial separations. This interaction progressively destabilizes the negatively polarized regions separating the domains, resulting in accelerated outward propagation of the central domain. Consequently, $R^{}_c$ exhibits a pronounced increment with $t$. In comparison, $R^{}_{\rm eq}$ increases more uniformly because it represents the cumulative contribution of all propagating domains. Increasing the applied electric field enhances the mobility of every domain wall, producing larger forward Type-III excitation during each pulse and significantly accelerating the growth of both $R^{}_c$ and $R^{}_{\rm eq}$.

The corresponding evolution with $n$ is shown in Fig.~\ref{Figure10}(c) and Fig.~\ref{Figure10}(d). The increase of $R^{}_c$ with $n$ clearly demonstrates the transition from independent domain growth to cooperative multi-domain propagation. Interestingly, the central domain initially undergoes accelerated growth until domain coalescence, after which the evolution is governed by the outward propagation of the merged outer boundary, resulting in a gradual decrease in the growth rate.
The behaviour of $R^{}_{\rm eq}$ reflects the collective growth of the entire multi-domain structure and therefore exhibits a smoother evolution. The impact of $T^{}_{\rm on}$ is illustrated in Fig.~\ref{Figure10}(e) and Fig.~\ref{Figure10}(h). Both $R^{}_c$ and $R^{}_{\rm eq}$ increase significantly with $T^{}_{\rm on}$. As the spacing between the domains is much smaller, it facilitates earlier interactions and coalescence, especially affecting $R^{}_c$ drastically. The increase in $R^{}_{\rm eq}$ reflects simultaneous growth of multiple-domains, indicating enhanced multi domain switching efficiency.

\begin{figure*}
	\includegraphics[scale=0.650]{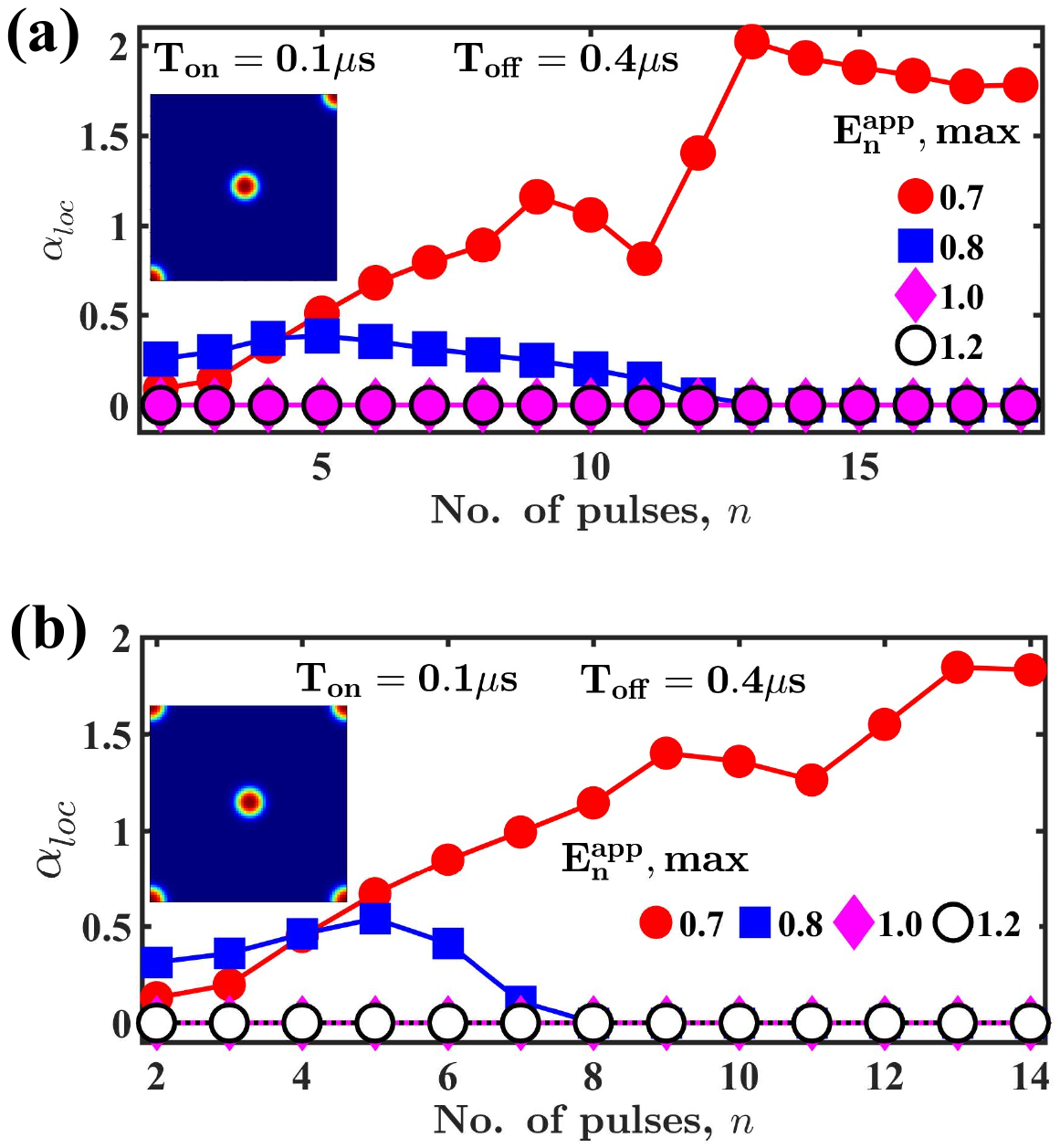}
	\caption{ Evolution of the local kinetic exponent $\alpha^{}_{\rm loc}$ in (a) Diagonal Corner–Centre Configuration (DCC) and (b) Corner–Centre Configuration (CCC) FE system. $\alpha^{}_{\rm loc}$ is plotted as a function of pulse number $n$ for different $E^{\rm app}_{n}, {\rm max}$ with fixed $T^{}_{\rm on}=0.1$ $\mu\mathrm{s}$ and $T^{}_{\mathrm{off}}=0.4$ $\mu\mathrm{s}$. Insets show the corresponding initial domain configurations. $\alpha_{\rm loc}>1$ corresponds to super-linear domain growth driven by cooperative field-assisted domain-wall propagation, $\alpha_{\rm loc}\approx1$ denotes a nearly self-similar propagation regime, and $\alpha^{}_{\rm loc}<1$ signifies decelerated growth arising from geometric confinement and relaxation effects.} 
	\label{Figure11}
\end{figure*}

To further quantify the influence of multi-domain interactions on the switching kinetics, we study the evolution of the local kinetic exponent $\alpha^{}_{\rm local}$ for the DCC and CCC configurations for various $E^{\rm app}_{n}, {\rm max}$ in Fig.~(\ref{Figure11}). We show the evolution of $\alpha^{}_{\rm local}$ in Fig.~\ref{Figure11}(a) and Fig.~\ref{Figure11}(b) for the DCC and CCC systems, respectively. Irrespective of $E^{\rm app}_{n}, {\rm max}$, both configurations exhibit three distinct characteristics, kinetic regimes during the switching process. During the initial pulse cycles, $\alpha^{}_{\rm local}$ increases rapidly, exceeding unity, indicating a super-linear acceleration regime. In such a scenario, successive voltage pulses progressively enhance the efficiency of irreversible Type-III excitation, allowing the domain walls to propagate a bit farther compared to the previous pulse cycle. On the other hand, $\alpha_{\rm local}$ remains close to unity because the driving force is only sufficient to sustain nearly constant domain-wall propagation at relatively low electric fields. As the electric field increases, there is a substantial increase in $\alpha^{}_{\rm local}$, indicating stronger pulse-to-pulse acceleration of the switching process. The DCC configuration exhibits a relatively broad acceleration regime because the larger separation between the central and corner domains allows the three domains to propagate independently over a greater number of pulse cycles before significant interaction occurs. Consequently, the increase in $\alpha^{}_{\rm local}$ is more gradual. In contrast, the CCC configuration exhibits a sharper rise in the local kinetic exponent followed by an earlier transition toward saturation. It is because the smaller separation between the central and corner domains promotes earlier cooperative interaction and domain coalescence, leading to more efficient Type-III excitation during the intermediate switching stage. Once coalescence is achieved, the remaining switching is governed primarily by the propagation of the common outer boundary, causing $\alpha^{}_{\rm local}$ to decrease more rapidly.
	
\begin{figure*}
	\includegraphics[scale=0.70]{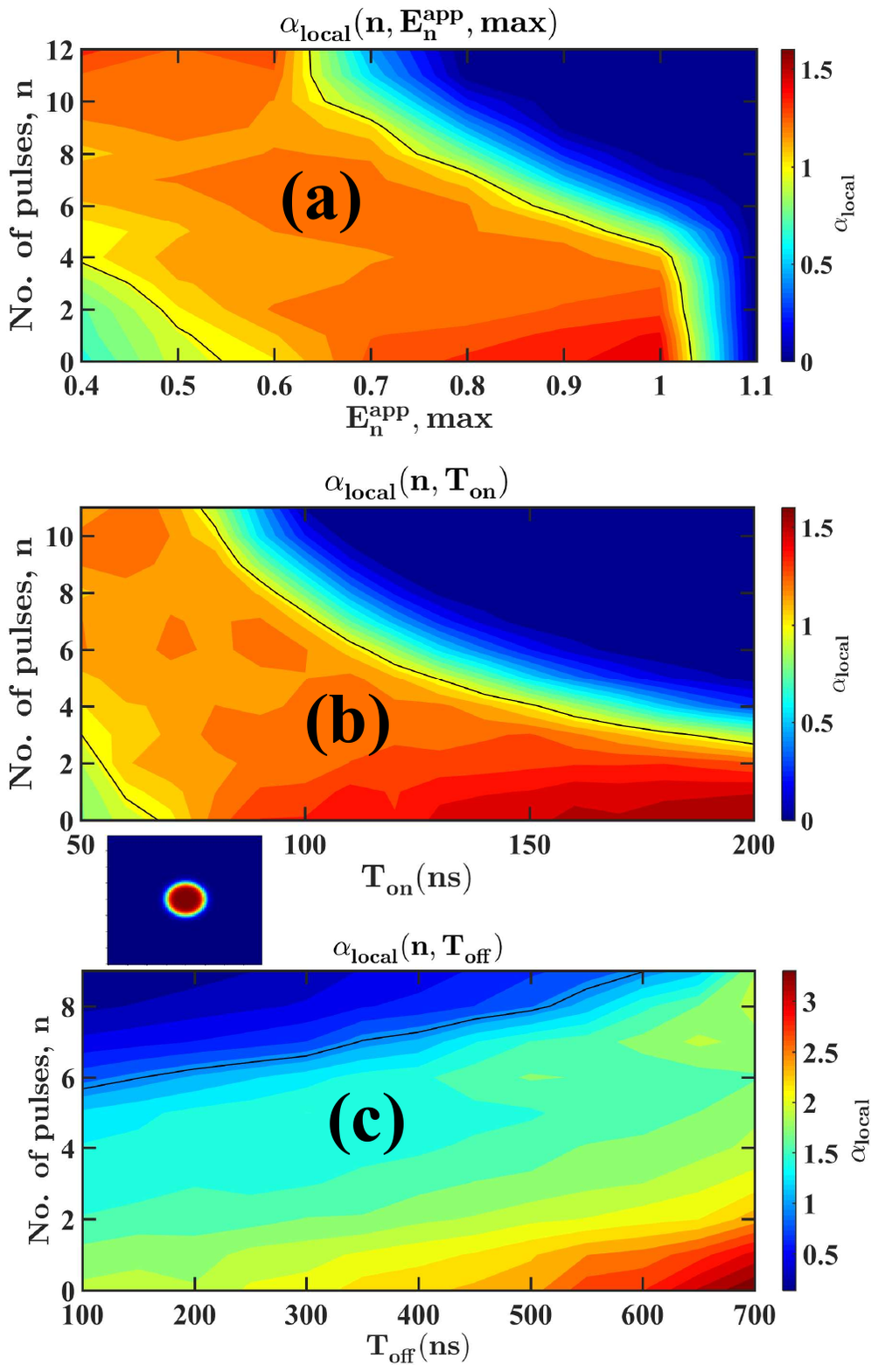}
	\caption{ Two-dimensional kinetic maps of $\alpha^{}_{\mathrm{local}}$ with centre domain (CeD) FE system. Contour map of $\alpha^{}_{\rm local}(n,E^{\rm app}_{n}, {\rm max})$ as a function of $n$ and $E^{\rm app}_{n}, {\rm max}$ for fixed $T_{\rm on}=0.1$ $\mu\mathrm{s}$ and $T^{}_{\rm off}=0.4$$\mu\mathrm{s}$. (b) Contour map of $\alpha^{}_{\rm local}(n,T^{}_{\rm on})$ for different $T^{}_{\rm on}$ with fixed $E^{\rm app}_{n}, {\rm max}=0.8$ and $T^{}_{\rm off}=0.4$ $\mu\mathrm{s}$. (c) Contour map of $\alpha^{}_{\rm local}(n,T_{\rm off})$ for different $T^{}_{\rm off}$ at fixed $E^{\rm app}_{n}, {\rm max}=0.8$ and $T^{}_{\rm on}=0.1$ $\mu\mathrm{s}$. $\alpha^{}_{\mathrm{local}}>1$ indicates accelerated growth, $\alpha^{}_{\mathrm{local}}\approx1$ self-similar domain-wall propagation, and $\alpha^{}_{\mathrm{local}}<1$ relaxation-dominated decelerating growth. The $\alpha^{}_{\mathrm{local}}\approx 1$ contour shifts to lower pulse numbers with increasing $E^{\rm app}_{n}, {\rm max}$ or $T^{}_{\rm on}$, but to higher pulse numbers with longer $T^{}_{\rm off}$, revealing transitions between excitation- and relaxation-controlled switching regimes.} 
	\label{Figure12}
\end{figure*}

Finally, to establish a comprehensive understanding of the switching kinetics under pulse-driven accumulative polarization, we study the evolution of $\alpha^{}_{\rm local}$ for CeD configurations over a broad range (very dense) of relevant parameters using two-dimensional phase maps in Fig.~(\ref{Figure12}). This analysis provides a unified description of how the instantaneous domain-growth kinetics evolve throughout the switching process and identifies the transition between different kinetic regimes under a wide range of operating conditions. Fig.~(\ref{Figure12})(a)-(c) 
shows the two-dimensional kinetic maps of the local growth exponent, $\alpha^{}_{\mathrm{local}}$, illustrating the evolution of domain-growth kinetics during pulse-driven accumulative polarization switching. 
 During the initial pulse cycles, $\alpha_{\rm local}$ increases rapidly and frequently exceeds unity with larger $E^{\rm app}_{n},\rm{max}$, indicating a super-linear acceleration regime. Physically, repeated voltage excitation progressively destabilizes the remaining negatively polarized regions through cumulative Type-I excitation while reinforcing the switched domains via repeated Type-II excitation, thereby promoting irreversible  domain-wall propagation and accelerating domain growth and polarization accumulation through Type-III excitation. Consequently, both the peak value of $\alpha^{}_{\rm local}$ and the duration over which $\alpha^{}_{\rm local}>1$ increase. 
 Similarly, increasing the pulse-on duration ($T^{}_{\rm on}$) extends the time over which the electric field acts on the domain wall, producing larger forward displacements and maintaining the acceleration regime over a wider range of $n$. Conversely, increasing the pulse-off duration weakens the acceleration regime because the domain wall experiences greater spontaneous backward motion before the arrival of the next pulse. Following the acceleration stage, $\alpha^{}_{\rm local}$ gradually approaches unity, corresponding to a steady propagation regime. In this regime, the enhancement of switching produced by repeated pulse excitation is balanced by the progressive reduction of the remaining switchable region, causing the domain wall to advance by nearly equal distances during successive pulse cycles. The accumulated polarization and equivalent domain radius therefore increase almost linearly with pulse number $n$. 

We now clearly state that the analyses presented in this work through Fig.~(\ref{Figure1})-Fig.~(\ref{Figure12}) demonstrate that pulse-driven accumulative switching in HZO is governed by the dynamic competition between Type-I, Type-II, and Type-III excitation and relaxation processes. While Type-I and Type-II mechanisms continuously modulate the local polarization within negatively and positively polarized regions, respectively, irreversible Type-III domain-wall propagation determines the net accumulation of polarization. The evolution of the local kinetic exponent $\alpha$ reveals that this competition proceeds through three universal kinetic regimes: acceleration, steady propagation, and deceleration, which can be systematically tuned by the initial domain configuration and the applied pulse parameters, viz. applied field pulse strength $E^{\rm app}_{n},\rm{max}$, pulse-on duration $T^{}_{\rm on}$ and pulse-off duration $T^{}_{\rm off}$. These observations establish a unified kinetic framework for understanding and optimizing accumulative switching in ferroelectric devices.	

\section{SUMMARY AND CONCLUSION}
In summary, we have developed a comprehensive kinetic framework for understanding external field  pulse-driven accumulation-mediated polarization switching in ferroelectric HZO by combining extensive phase-field simulations with comprehensive analysis of domain evolution through dynamic scaling routes. Unlike previous studies that primarily focused on the physical origin of polarization accumulation, the present work establishes the dynamic regimes and scaling laws governing nonequilibrium switching under sequential sub-coercive electric-field pulses. Through rigorous computational and analytical analysis, we have successfully demonstrated that the macroscopic evolution of polarization is dictated by microscopic domain-wall dynamics arising from the competition between field-driven excitation during the pulse-on interval and spontaneous relaxation during the pulse-off interval. The latter is found to be strongly dependent on initial domain configurations, applied field pulse amplitude $E^{\rm app}_{n},\rm{max}$, pulse-on duration $T^{}_{\rm on}$ and pulse-off duration $T^{}_{\rm off}$.

A central and one of the most important outcomes of the present work is the identification of three distinct kinetic regimes governing accumulative switching. During the initial stage, successive voltage pulses progressively enhance irreversible domain-wall propagation, resulting in an acceleration regime characterized by super-linear domain growth characterized by a local kinetic exponent $\alpha_{\mathrm{local}}>1$. As switching progresses, the system evolves into a steady-propagation regime where $\alpha^{}_{\mathrm{local}}\approx1$, indicating nearly self-similar domain expansion with successive pulses. At later stages, the depletion of the remaining switchable polarization, increasing geometric constraints, and relaxation-driven backward domain-wall motion collectively suppress the switching kinetics, giving rise to a deceleration regime with $\alpha_{\mathrm{local}}<1$. These kinetic transitions constitute a universal signature of pulse-driven accumulative switching and provide a quantitative description of the evolution of domain-growth dynamics throughout the switching process.

The present study further demonstrates that the initial domain configuration plays a decisive role in determining the polarization switching pathway by controlling the available domain-wall length, propagation symmetry, and cooperative interactions between neighbouring domains. Consequently, domain configurations possessing unrestricted propagation pathways (isotropical domain configurations) exhibit faster polarization accumulation and larger transient values of the local kinetic exponent $\alpha^{}_{\mathrm{local}}$ than geometrically constrained systems. Likewise, increasing the electric-field amplitude $E^{\rm app}_{n},\rm{max}$, and pulse-on duration $T^{}_{\rm on}$ strengthens irreversible domain-wall propagation dominated through Type-III-excitation, thereby extending the acceleration regime and significantly reducing the number of pulses required for complete switching. On the contrary, longer pulse-off durations $T^{}_{\rm off}$ enhance spontaneous relaxation, weaken the cumulative effect of successive pulses, and delay the onset of rapid domain growth. So, these observations establish a direct quantitative relationship between pulse parameters, domain-wall kinetics, and polarization accumulation.


In conclusion, the present work establishes a unified microscopic picture of pulse-driven accumulative switching in ferroelectric HZO, demonstrating that the nonequilibrium switching dynamics obey well-defined kinetic scaling laws governed by the interplay between excitation and relaxation processes. Beyond providing fundamental insight into ferroelectric domain-wall dynamics, our proposed framework offers predictive guidelines for engineering pulse protocols that precisely control intermediate polarization states, thereby facilitating the development of energy-efficient ferroelectric memories, multilevel storage technologies, and neuromorphic computing architectures based on accumulative polarization switching.

\section*{Conflict of Interest}
The authors declare no conflicts of interest.

\section*{DATA AVAILABILITY}
The data that support the findings of this study are available from the corresponding author upon reasonable request.
\bibliography{ref}
\end{document}